\documentclass[twocolumn, twocolappendix]{aastex701}

\usepackage{amsmath}
\usepackage{color}
\usepackage{comment}

\begin{document}

\title{Dense Ionized Outflow with Five Narrow Components in a Dust-obscured Galaxy}

%\correspondingauthor{August Muench}
%\email{greg.schwarz@aas.org, gus.muench@aas.org}

\author[0009-0001-9947-6732]{Taketo Yoshida}
\affiliation{Graduate School of Science and Engineering, Ehime University, 2-5 Bunkyo-cho, Matsuyama, Ehime 790-8577, Japan}
\email[show]{yoshida@cosmos.phys.sci.ehime-u.ac.jp}

\author[0000-0002-7402-5441]{Tohru Nagao}
\affiliation{Research Center for Space and Cosmic Evolution, Ehime University, 2-5 Bunkyo-cho, Matsuyama, Ehime 790-8577, Japan}
\affiliation{Amanogawa Galaxy Astronomy Research Center, Kagoshima University, 1-21-35 Korimoto, Kagoshima 890-0065, Japan}
\email{tohru.nagao@gmail.com}

\author[0000-0002-3531-7863]{Yoshiki Toba}
\affiliation{Department of Physical Sciences, Ritsumeikan University, 1-1-1, 
Noji-higashi, Kusatsu, Shiga 525-8577, Japan}
\affiliation{Academia Sinica Institute of Astronomy and Astrophysics, 11F of Astronomy-Mathematics Building, AS/NTU, No.1, Section 4, Roosevelt Road, Taipei 10617, Taiwan}
\affiliation{Research Center for Space and Cosmic Evolution, Ehime University, 2-5 Bunkyo-cho, Matsuyama, Ehime 790-8577, Japan}
\email{toba@fc.ritsumei.ac.jp}

\author[0009-0008-8432-7460]{Naomichi Yutani}
\affiliation{Department of Planetology, Graduate School of Science, Kobe University, 1-1 Rokkodai, Nada-ku, Kobe, Hyogo 657-8501, Japan}
\email{yutaninm@gmail.com}

\author[0000-0002-4375-254X]{Thomas Bohn}
\affiliation{Research Center for Space and Cosmic Evolution, Ehime University, 2-5 Bunkyo-cho, Matsuyama, Ehime 790-8577, Japan}
\email{bohn.thomas_carl.gb@ehime-u.ac.jp }

\author[0009-0001-8283-6308]{Kohei Shibata}
\affiliation{Graduate School of Science and Engineering, Ehime University, 2-5 Bunkyo-cho, Matsuyama, Ehime 790-8577, Japan}
\email{shibata@cosmos.phys.sci.ehime-u.ac.jp}

\author[0009-0008-6886-9315]{Nozomu Tamada}
\affiliation{Graduate School of Science and Engineering, Ehime University, 2-5 Bunkyo-cho, Matsuyama, Ehime 790-8577, Japan}
\email{nozomu.tamada@gmail.com}

\begin{abstract}

We present our discovery of a complex ionized outflow in SDSS J101034.28+372514.7 (J1010+3725), a dust-obscured galaxy (DOG) at $z=0.282$. 
The SDSS optical spectrum of J1010+3725 shows five narrow components with one broad component in [O{\,\sc iii}]$\lambda$5007, which represents one of the most complex outflow structures observed among dusty active galactic nuclei.
Spectrum fitting shows that the five narrow components have a wide range of velocity shifts (from $-1475$ to $+507$ km s$^{-1}$). 
The possible multiple peaks are also observed in [O{\,\sc iii}]$\lambda$4363 and [Ne{\,\sc iii}]$\lambda$3868, which allows us to investigate the physical condition of the outflowing gas by comparing the measured emission-line flux ratios with photoionization models. 
The comparison suggests that the five outflowing components are characterized by high hydrogen densities ($\gtrsim 10^5$ cm$^{-3}$). 
Our results imply that the five highly dense gas components may be outflowing with multiple bulk velocities at the innermost part of the narrow-line region in J1010+3725. 

\end{abstract}

%% Keywords should appear after the \end{abstract} command. 
%% The AAS Journals now uses Unified Astronomy Thesaurus concepts:
%% https://astrothesaurus.org
%% You will be asked to selected these concepts during the submission process
%% but this old "keyword" functionality is maintained in case authors want
%% to include these concepts in their preprints.

\section{Introduction} \label{sec:intro}

It has been shown that the mass of supermassive black holes (SMBHs) strongly correlates with their host galaxy properties such as velocity dispersion and bulge mass, suggesting that they have co-evolved (e.g., Magorrian et al. \citeyear{Magorrian+98}; Marconi \& Hunt \citeyear{Marconi+03}; Kormendy \& Ho \citeyear{Kormendy+13}).
A major merger of gas-rich galaxies is often assumed to explain this co-evolution (e.g., Sanders et al. \citeyear{Sanders+88}; Hopkins et al. \citeyear{Hopkins+08}). 
In this scenario, the merging/interaction of gas-rich galaxies induces a burst of
star-formation (SF) and the infall of gas onto the nucleus (e.g., Barnes and Hermquist \citeyear{Barnes+91}), the latter of which ignites the active galactic nucleus (AGN) at the center of the galaxy (hereafter the major merger scenario). 
During this phase, the star forming regions and AGN are typically enshrouded by a large amount of dust. 
Eventually, the surrounding gas and dust are blown out by powerful radiation and/or outflow from the AGN.

Numerical simulations have suggested that gas-rich major mergers of galaxies can be classified as dust-obscured galaxies (DOGs: Dey et al. \citeyear{Dey+08}) around the peak of their SF rate and AGN accretion (Narayanan et al. \citeyear{Narayanan+10}; Yutani et al. \citeyear{Yutani+22}). 
DOGs are selected by their extreme optical to infrared (IR) color ($i - [22] \geq 7$: Toba et al. \citeyear{Toba+15}; see also Dey et al. \citeyear{Dey+08}), and include both dusty SF galaxies and dusty AGNs. 
These two populations of DOGs can be separated by their characteristic features in their spectral energy distributions (SEDs: Dey et al. \citeyear{Dey+08}; Melbourne et al. \citeyear{Melbourne+12}; Toba et al. \citeyear{Toba+15}; Noboriguchi et al. \citeyear{Nobo+19}; Suleiman et al. \citeyear{Suleiman+22}; Yu et al. \citeyear{Yu+24}; Yoshida et al. \citeyear{Yoshida+25}). 
SF-dominated DOGs are often called ``bump DOGs'' owing to their characteristic bump feature in their SEDs at near-infrared (NIR) wavelength (e.g., Desai et al. \citeyear{Desai+09}; Melbourne et al. \citeyear{Melbourne+12}; Yu et al. \citeyear{Yu+24}). 
On the contrary, AGN-dominated DOGs are characterized by power-law SED shapes from optical to mid-IR (MIR) wavelengths (e.g., Bussmann et al. \citeyear{Bussmann+09}; Melbourne et al. \citeyear{Melbourne+12}; Zou et al. \citeyear{Zou+20}; Yu et al. \citeyear{Yu+24}), and are thus called ``power-law (PL) DOGs''.

Some AGN-dominated DOGs are characterized by their hotter dust emission from relatively strong AGN activity ($\gtrsim$ 60K; Wu et al. \citeyear{Wu+12}; Fan et al. \citeyear{Fan+17}), and they are known as Hot DOGs (Eisenhardt et al. \citeyear{Eisenhardt+12}; Wu et al. \citeyear{Wu+12}).
They are characterized by higher IR luminosities and obscuration than those of classical DOGs (e.g., Tsai et al. \citeyear{Tsai+15}; Toba et al. \citeyear{Toba+18}, \citeyear{Toba+20}; Li et al. \citeyear{Li+24}).
Based on the major merger scenario, dusty AGNs including PL DOGs and Hot DOGs are expected to host strong outflows, which have the potential to eject gas and dust to the outer regions of the galaxy.
In fact, both observations and simulations suggest that these dusty AGNs often show complex [O{\,\sc iii}] emission profiles, which are more significant in more heavily obscured AGNs (e.g, Zakamska et al. \citeyear{Zakamska+16}; Toba et al. \citeyear{Toba+17}; Perrotta et al. \citeyear{Perrotta+19}; Finnerty et al. \citeyear{Finnerty+20}; Jun et al. \citeyear{Jun+20}; Yutani et al. \citeyear{Yutani+24}).

Yoshida et al. (in preparation) identified DOGs having both a Sloan Digital Sky Survey (SDSS: York et al. \citeyear{York+00}) spectrum and a NIR counterpart to investigate the relation between their SED shape and the properties of their SDSS spectrum. 
Within this sample, we report the serendipitous discovery of multi-peaked [O{\,\sc iii}]$\lambda$$\lambda$4959, 5007  emission profiles, as well as possible multiple peaks in [Ne{\,\sc iii}]$\lambda$3868 and [O{\,\sc iii}]$\lambda$4363, in SDSS J101034.28+372514.7 (J1010+3725), indicative of a multi-component outflow. 
In this paper, we report the unique nature and possible implications of this extraordinary outflow feature found in the spectrum of a luminous DOG. 

This paper is organized as follows: Section \ref{sec: data} presents the photometric and spectroscopic datasets of J1010+3725.
We show the SED and spectrum of J1010+3725, and perform spectrum fitting in Section \ref{sec:results}.
In Section \ref{sec:discussion}, we discuss the nature of J1010+3725 based on the comparison of its line properties with those of other AGN classes and a theoretical model. 
%The cosmology adopted in this paper assumes a flat universe with $H_{0} = 70$ kms$^{-1}$ Mpc$^{-1}$, $\Omega _{M} = 0.3$, and $\Omega _{\Lambda} =0.7$. 
Unless otherwise noted, all magnitudes refer to the AB system.

\section{Data} \label{sec: data}
J1010+3725 has multi-wavelength data from ultraviolet (UV) to MIR.
We obtained UV, optical, NIR, and MIR data from the {\it Galaxy Evolution Explorer} ($GALEX$: Martin et al. \citeyear{Martin+05})
%% ==========
%%  footnote 
%% ==========
\footnote{The $GALEX$ data used in this paper can be found in MAST: \cite{STScI}}
%(DOI: \dataset[10.17909/T9H59D]{http://dx.doi.org/10.17909/T9H59D})}
%% =====
%%  end
%% =====
, SDSS, Two Micron All Sky Survey (2MASS: Skrutskie et al. \citeyear{Skrutskie+06}), and {\it Wide-field Infrared Survey Explorer} ({\it WISE}; Wright et al. \citeyear{Wright+10}), respectively. 
To match the counterparts among the datasets, we first cross-matched the SDSS position of J1010+3725 with the AllWISE catalog (Wright et al. \citeyear{Wright+19}; Cutri et al. \citeyear{Cutri+21}) by a search radius of 3$^{\prime\prime}$ (Toba \& Nagao \citeyear{Toba+16}) and referred to the 2MASS photometry in this AllWISE catalog as the NIR data. 
This object has an extremely red color in its optical-to-MIR range ($i - [22] = 7.23 \pm 0.03$) and thus satisfies the DOG criterion ($i - [22] \geq 7$).
Although this object was previously included in the DOG sample of Toba \& Nagao (\citeyear{Toba+16}; see also Toba et al. \citeyear{Toba+17}; Zou et al. \citeyear{Zou+20}), its complex emission-line profile has not yet been investigated in detail.
We also performed cross-matching with $GALEX$ and employed a 3$^{\prime\prime}$ search radius (Toba et al. \citeyear{Toba+22}, \citeyear{Toba+24}).
Note that J1010+3725 does not have any far-infrared (FIR) or radio counterparts within the Herschel Space  Observatory (Pilbratt et al. \citeyear{Pilbratt+10}), {\it AKARI} (Murakami et al. \citeyear{Murakami+07}), Infrared Astronomical Satellite (Moshir et al. \citeyear{Moshir+92}), or the Very Large Array Faint Images of the Radio Sky at Twenty-Centimeters survey (Becker et al. \citeyear{Becker+95}). 
The basic information and photometries are summarized in Table \ref{tab: info}.

%% =======
%%  table 
%% =======
\begin{table}[t]
\caption{Observed Properties
\label{tab: info}}
\centering
\begin{tabular}{lc}
\hline
\hline
J101034.28+372514.7  \\
\hline
R.A. (SDSS) [J2000.0] & 10:10:34.28 \\
Decl. (SDSS) [J2000.0] & +37:25:14.76 \\
Redshift (SDSS) & 0.282 \\ %$0.282302 \pm 0.000034$
$GALEX$ $FUV$ [$\mu$Jy] & 12.56 $\pm$ 4.05 \\
$GALEX$ $NUV$ [$\mu$Jy] & 19.70 $\pm$ 3.88 \\
SDSS $u$ [$\mu$Jy] & 69.09 $\pm$ 1.51 \\
SDSS $g$ [$\mu$Jy] & 89.26 $\pm$ 0.78 \\
SDSS $r$ [$\mu$Jy] & 186.67 $\pm$ 1.26 \\
SDSS $i$ [$\mu$Jy] & 207.19 $\pm$ 1.58 \\
SDSS $z$ [$\mu$Jy] & 374.18 $\pm$ 4.64 \\
2MASS $J$ [mJy] & 0.55 $\pm$ 0.04 \\
2MASS $H$ [mJy] & 1.28 $\pm$ 0.06 \\
2MASS $K_{\rm s}$ [mJy] & 2.11 $\pm$ 0.07 \\
{\it WISE} $W1$ [mJy] & 4.44 $\pm$ 0.09 \\
{\it WISE} $W2$ [mJy] & 8.26 $\pm$ 0.15 \\
{\it WISE} $W3$ [mJy] & 38.12 $\pm$ 0.56 \\
{\it WISE} $W4$ [mJy] & 161.88 $\pm$ 3.58 \\
\hline
\end{tabular}
\end{table}
%% =====
%%  end
%% =====

Based on the SDSS spectrum (Plate-MJD-FiberID: 1426-52993-0110; on-source integration time: 8100 sec), J1010+3725 has a spectroscopic redshift of $z = 0.282$. Further details of this spectrum are presented in Section \ref{subsec: spec}.

%% ========
%%  figure 
%% ========
\begin{figure*}[t]
\includegraphics[width=17.7cm]{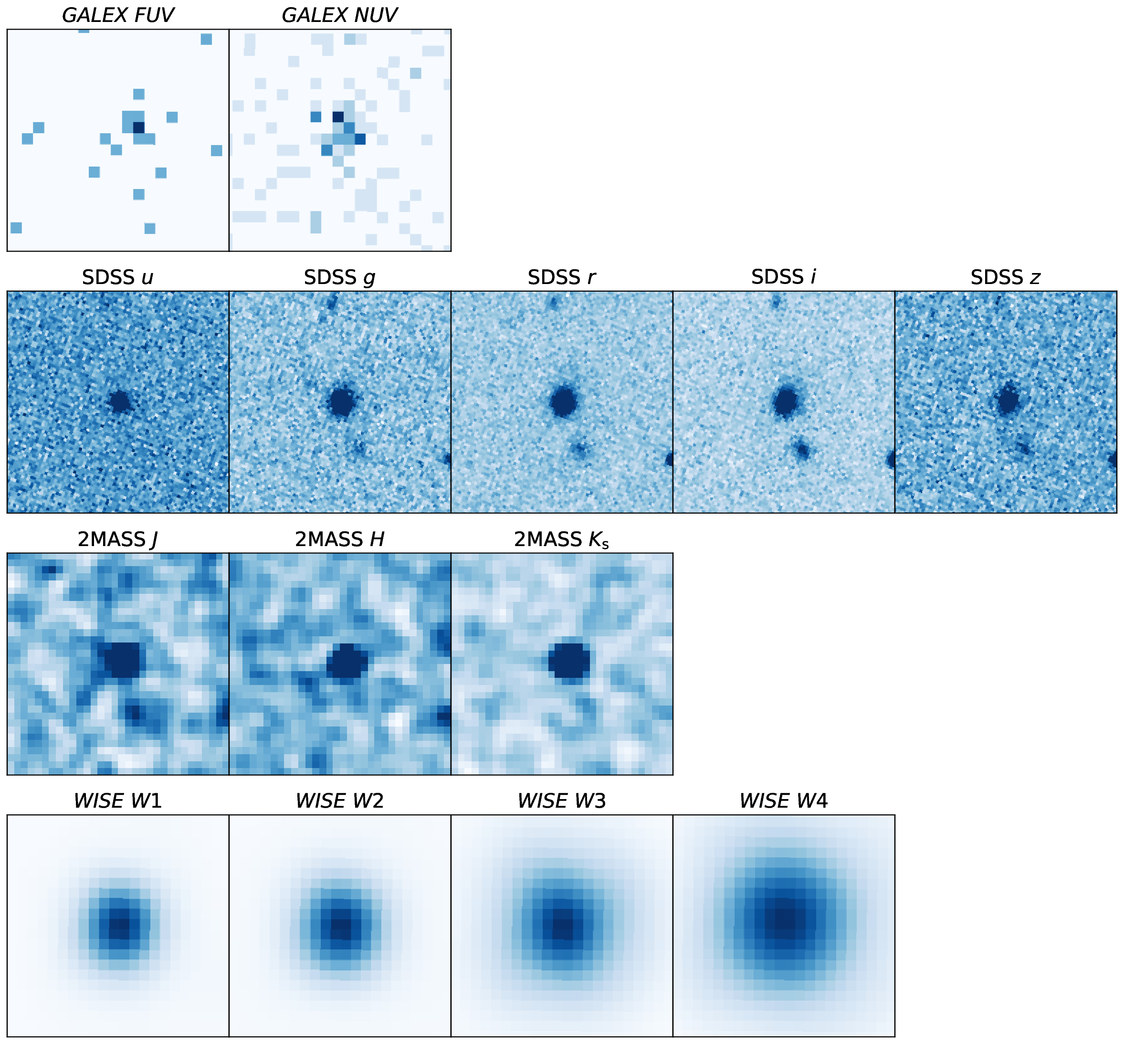}
\caption{
Thumbnail images of J1010+3725 with a size of $30^{\prime\prime} \times 30^{\prime\prime}$. 
$GALEX$ ($FUV$ and $NUV$), SDSS ($u$, $g$, $r$, $i$, and $z$), 2MASS ($J$, $H$, and $K_{\rm s}$), and {\it WISE} ($W1$, $W2$, $W3$, and $W4$) images are shown from top to bottom. 
North is up and east is to the left. 
\label{fig: image}}
\end{figure*}
%% =====
%%  end
%% =====

\section{Results} \label{sec:results}

\subsection{Spectral Energy Distribution} \label{subsec: SED}

%% ========
%%  figure 
%% ========
\begin{figure*}[t]
\includegraphics[width=17.7cm]{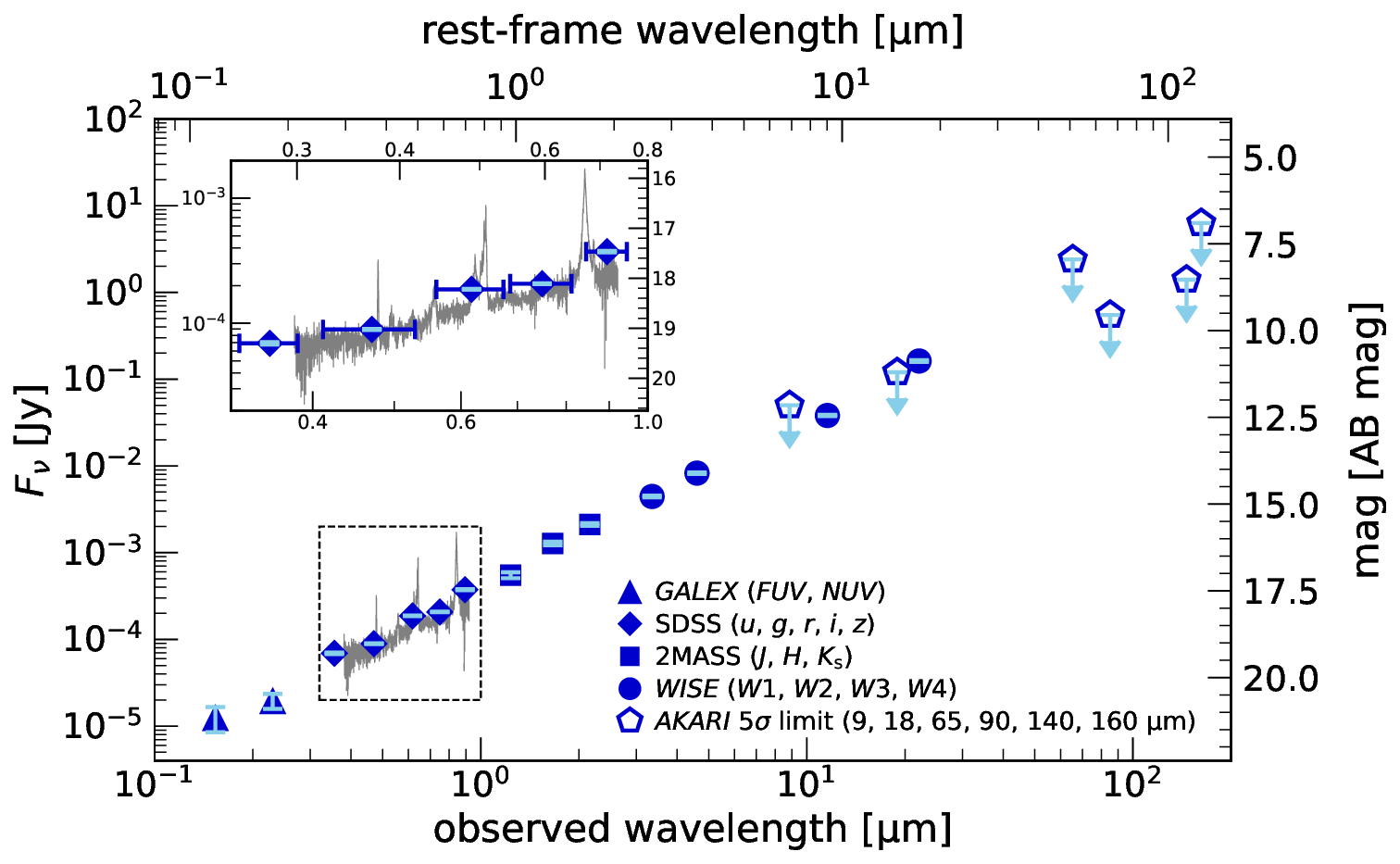}
\caption{
SED and spectrum of J1010+3725 ($z=0.282$). %$0.282302 \pm 0.000034$
The blue filled triangles, diamonds, squares, and circles represent $GALEX$, SDSS, 2MASS, {\it WISE} photometries, respectively. 
The blue opened pentagons denote the 5$\sigma$ detection limits of {\it AKARI}. 
The gray line depicts the SDSS spectrum. 
The top-left inset shows a zoom-in view of the SDSS photometry and spectrum. 
\label{fig: SED}}
\end{figure*}
%% =====
%%  end
%% =====

Figure \ref{fig: image} shows the multiwavelength images of J1010+3725. 
One faint object appears in the south direction from J1010+3725 with a separation
of 6.67 arcsec. 
This object is seen in the images from SDSS $g$-band to 2MASS $J$-band but disappears in the 2MASS $H$-band and longer wavelength images, suggesting that the color of this object is relatively blue. 
Therefore, plausibly, this blue object does not contaminate the AllWISE photometry given the low resolution of the {\it WISE} images.

Figure \ref{fig: SED} shows the SED of J1010+3725 from NUV to FIR (FIR data are the 5$\sigma$ detection limit of {\it AKARI}). 
The SED clearly follows a power law and is thus classified as PL DOG, meaning that this object is plausibly an AGN-dominated DOG. 
This result is consistent with some AGN features observed in the SDSS spectrum (see Section \ref{subsec: spec}).
Recently, Yoshida et al. (\citeyear{Yoshida+25}) suggested that PL DOGs in early evolutionary stages of the major merger scenario show a break feature in their SED because their heavier obscuration causes a steeper slope at optical-NIR than that observed in late stages. 
However, such a feature is not detected, suggesting that J1010+3725 is in a late phase of galaxy evolution and is expected to be blowing out the surrounding dust.

\subsection{Spectral Properties} \label{subsec: spec}

%% ========
%%  figure 
%% ========
\begin{figure*}[t]
\includegraphics[width=17.7cm]{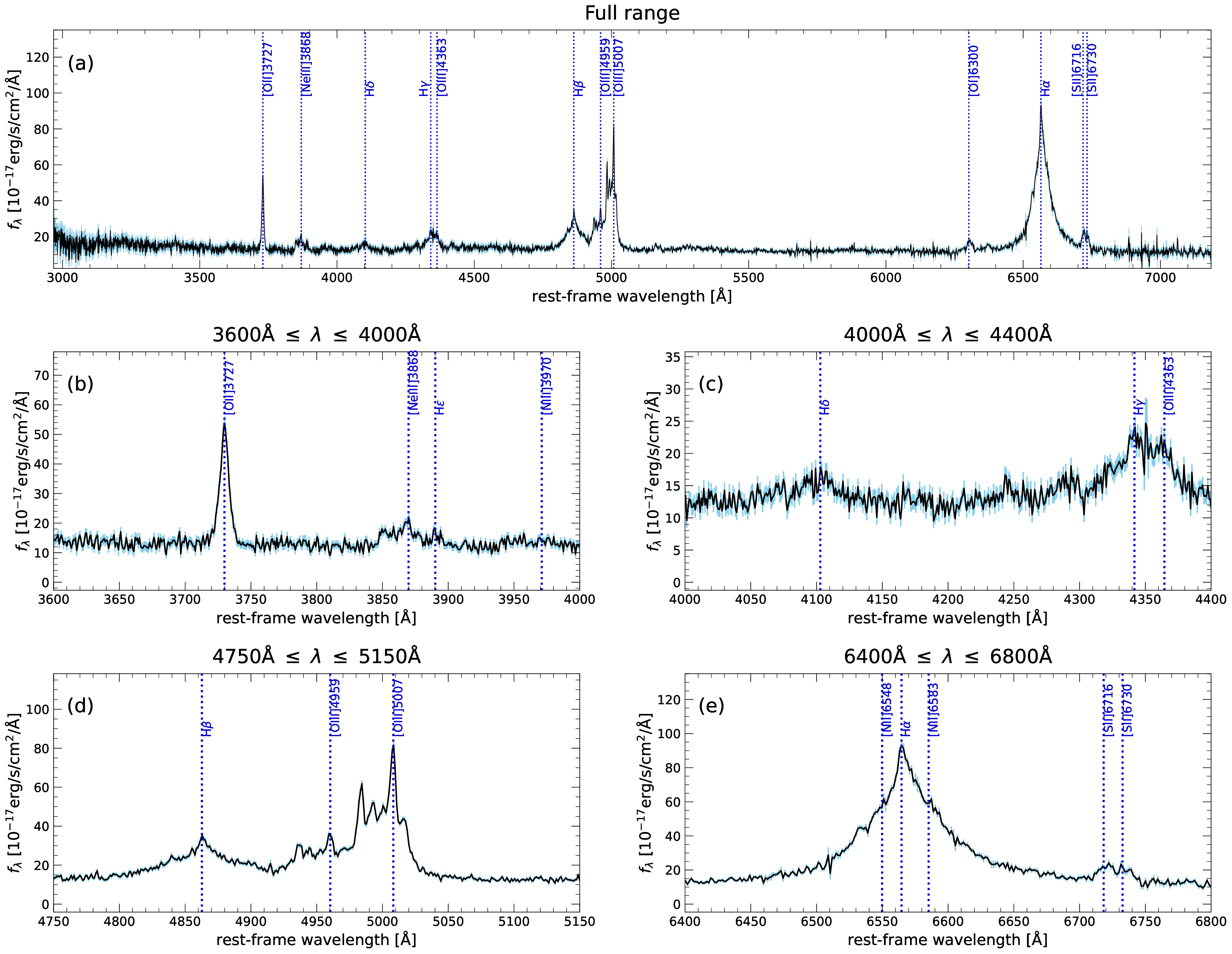}
\caption{
Optical spectrum (black solid line) and its error (light-blue bar) of J1010+3725 obtained from SDSS. Panel (a) shows the full rest-frame wavelength coverage, whereas panels (b)–(e) present enlarged views around selected emission lines: (b) [Ne{\,\sc iii}]$\lambda$3868, (c) [O{\,\sc iii}]$\lambda$4363, (d) [O{\,\sc iii}]$\lambda$$\lambda$4959, 5007 and H$\beta$, and (e) H$\alpha$. 
\label{fig: spec}}
\end{figure*}
%% =====
%%  end
%% =====

Figure \ref{fig: spec} shows the rest-frame SDSS spectra of J1010+3725. 
The Balmer lines show broad components, indicative of the presence of an AGN in this system. 
This is consistent with the characteristic power-law SED observed in this galaxy (see Section \ref{subsec: SED}).
In Figure \ref{fig: spec} (d), five narrow peaks are clearly observed for both [O{\,\sc iii}]$\lambda$$\lambda$4959, 5007 emission profiles, each consisting of one rest component (most significant), one redshifted component, and three blueshifted components. 
Multi-peak emission lines have been previously observed in some other AGNs. 
However, most of them have double-peak emission lines, which are usually explained not only by an extreme outflow kinematics but also by binary (merging) AGN systems (e.g., Wang et al. \citeyear{Wang+09}; Liu et al. \citeyear{Liu+10}). 
Emission lines with more than two peaks cannot be explained by binary AGN scenario; therefore, the five-peak emission line in J1010+3725 allows us to carry out a more direct analysis of extreme emission-line kinematics without ambiguity from binary AGN interpretations.  
Furthermore, the five-peak emission-line profile represents an extreme case of complex kinematics compared with dusty AGNs having multi-peak emission lines (e.g., IRAS F14394+5332E; Rodr\'{i}guez Zaur\'{i}n et al. \citeyear{Rod+13}; Spence et al. \citeyear{Spence+18}; see also Tadhunter et al. \citeyear{Tadhunter+18}, \citeyear{Tadhunter+19}).  
Interestingly, [Ne{\,\sc iii}]$\lambda$3868 and [O{\,\sc iii}]$\lambda$4363 also show potentially multiple peaks (see Figure \ref{fig: spec} (b) and (c)), allowing us to investigate the condition of the outflow component based on the emission-line flux ratios (see Section \ref{subsec: cloudy}). 
Note that multiple peaks are not clearly observed in the [O{\,\sc iii}]$\lambda$4363 emission owing to the low signal-to-noise ratios (S/N) and the blend with the neighboring H$\gamma$ emission (see Section \ref{subsec: fit} for more details). 
The low-ionized [O{\,\sc ii}]$\lambda$$\lambda$3727, 3729 doublet line does not show such a complex velocity structure (see Figure \ref{fig: spec} (b)). 
This suggests a highly dense or highly ionized gas nature of the outflow components (see Section \ref{subsec: cloudy}).

\subsection{Spectrum Fitting} \label{subsec: fit}

%% ========
%%  figure 
%% ========
\begin{figure*}[t]
\includegraphics[width=17.7cm]{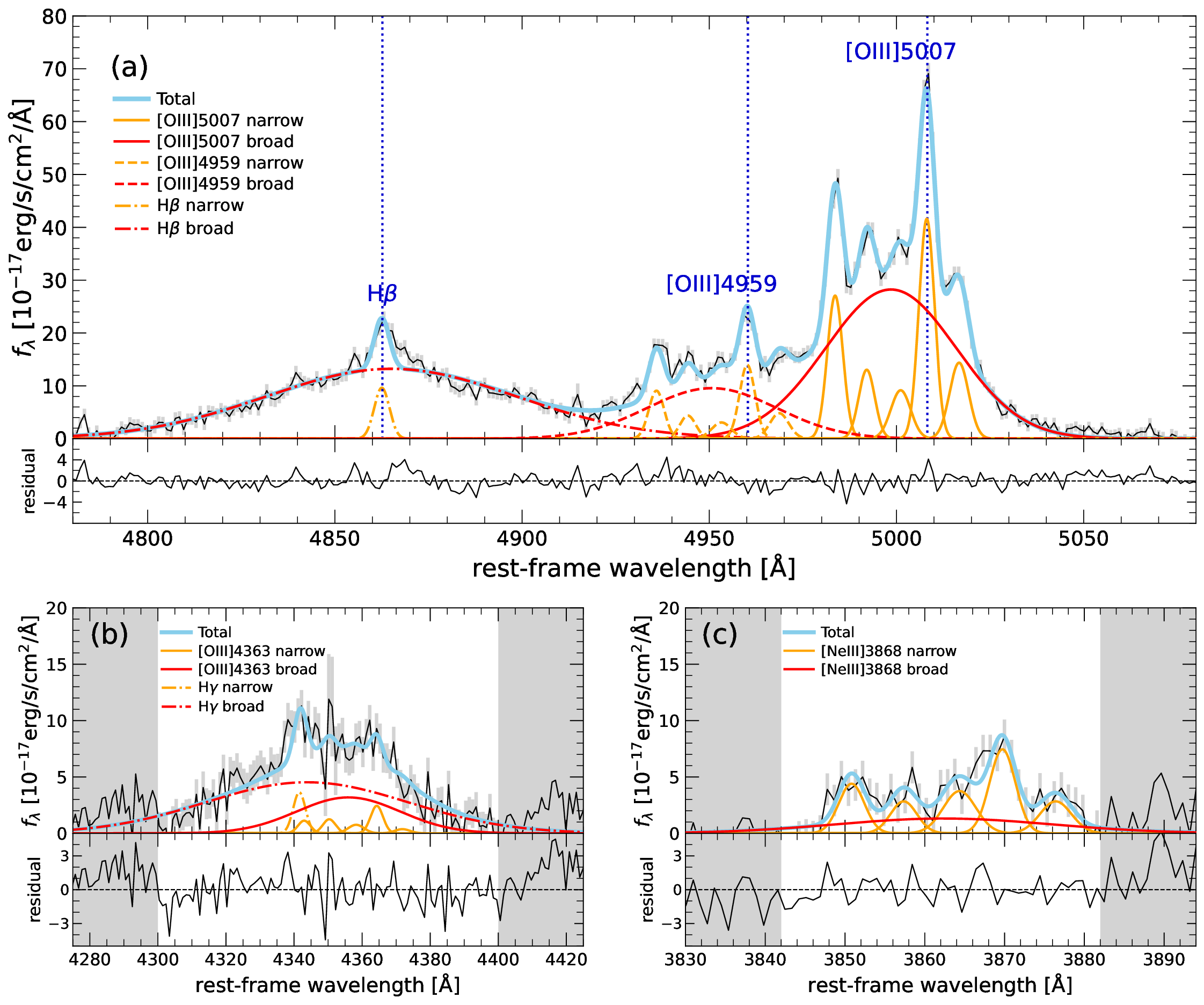}
\caption{
Continuum-subtracted spectrum with the results of spectrum fitting for (a) [O{\,\sc iii}]$\lambda$$\lambda$4959, 5007 and H$\beta$, (b) [O{\,\sc iii}]$\lambda$4363 and H$\gamma$, and (c) [Ne{\,\sc iii}]$\lambda$3868. 
The bold light-blue line represents the best-fit model. 
The orange and red lines correspond to the narrow and broad components for each line (see the legends), respectively. 
The vertical blue dotted line in panel (a) denotes the rest-frame wavelengths of each emission line. 
In panels (b) and (c), the spectra in the gray shade regions were excluded from the fitting to avoid the effects from other lines.  
The bottom subpanels show the residuals. 
\label{fig: fit}}
\end{figure*}
%% =====
%%  end
%% =====

First, we fitted the continuum with a power law. %$f _{\lambda} \propto \lambda ^{\alpha}$
To perform the continuum fitting, we excluded wavelength ranges around the most prominent emission lines or regions dominated by noise. %i.e., the edges of the spectrum
The resulting reduced $\chi ^2$ was $1.29$. 
This fitted power-law continuum was subtracted from the spectrum before fitting each line.

Second, we performed spectrum fitting for the [O{\,\sc iii}]$\lambda$$\lambda$4959, 5007 and H$\beta$ lines simultaneously.
Here, we assumed that [O{\,\sc iii}]$\lambda$5007 consists of five narrow components and one broad component. 
We set the amplitude of the Gaussian, velocity shift, and velocity width as free parameters for each component. 
We tied the velocity shift and width of the six [O{\,\sc iii}]$\lambda$4959 components (five narrow and one broad) to those of [O{\,\sc iii}]$\lambda$5007, while keeping the flux ratios between [O{\,\sc iii}]$\lambda$5007 and [O{\,\sc iii}]$\lambda$4959 fixed to 2.98 (Storey and Zeippen \citeyear{Storey+00}).
Therefore, the [O{\,\sc iii}]$\lambda$4959 components did not require any additional free parameters. 
For the H$\beta$ line, we assumed one narrow and one broad component, and set its narrow component to have the same kinematics as the most significant [O{\,\sc iii}] narrow component. 
Based on this model, we run a Markov chain Monte Carlo (MCMC) algorithm with $10^5$ steps to fit the spectrum using the Python package \texttt{emcee} (Foreman--Mackey et al. \citeyear{Foreman-Mackey+13}). 
After discarding the first 5000 steps of the MCMC chain to remove the burn-in phase, we adopted the median values and the 16th--84th percentile of the chain as the best-fit model and its uncertainty, respectively. 
Figure \ref{fig: fit} (a) shows the best-fit model. 
The reduced $\chi ^2$ of the best-fit model is 1.56.

Third, we also fitted [O{\,\sc iii}]$\lambda$4363 (with H$\gamma$) and [Ne{\,\sc iii}]$\lambda$3868 separately.
We assumed that all six components of [O{\,\sc iii}]$\lambda$4363 and [Ne{\,\sc iii}]$\lambda$3868 shared the same kinematics as [O{\,\sc iii}]$\lambda$5007, and that H$\gamma$ also shared the same kinematics as H$\beta$. 
Therefore, we set the amplitudes of emission-line components as the only free parameters. 
The best-fit models for [O{\,\sc iii}]$\lambda$4363 and [Ne{\,\sc iii}]$\lambda$3868 are shown in Figure \ref{fig: fit} (b) and (c), respectively. 
We obtained the reduced $\chi ^2$ values of 1.22 and 0.75 for [O{\,\sc iii}]$\lambda$4363 (with H$\gamma$) and [Ne{\,\sc iii}]$\lambda$3868, respectively. 
Although the reduced $\chi ^2$ value for the overall [O{\,\sc iii}]$\lambda$4363 fitting (with H$\gamma$) is good, three blueshift components and one redshift component have large uncertainties (S/N $\sim$ 1) due to the significant blending among some components and their week fluxes compared with noise. 
However, our conclusion remains unchanged within these large uncertainties (see Section \ref{subsec: cloudy}).

\subsection{Velocity Shifts and Velocity Dispersions} \label{subsec: velo}

%% =======
%%  table 
%% =======
\begin{table}[t]
\caption{[O{\,\sc iii}] kinematics
\label{tab: velo}}
\centering
\begin{tabular}{lcc}
\hline
\hline
component &  velocity shift & velocity dispersion\tablenotemark{a} \\
 & [km s$^{-1}$] & [km s$^{-1}$] \\
\hline
blueshift1 & $-1475^{+5}_{-5}$ & $107^{+10}_{-9}$ \\
blueshift2 & $-969^{+12}_{-12}$ & $112^{+17}_{-16}$ \\
blueshift3 & $-426^{+50}_{-28}$ & $145^{+63}_{-43}$ \\
rest & $-9^{+7}_{-6}$ & $108^{+8}_{-8}$ \\
redshift & $508^{+10}_{-11}$ & $138^{+16}_{-14}$ \\
broad & $-582^{+20}_{-21}$ & $1046^{+36}_{-29}$ \\
\hline
\end{tabular}
\tablenotetext{a}{Corrected for the instrumental broadening.}
\end{table}
%% =====
%%  end
%% =====

We derived the velocity shifts ($v$) and velocity dispersions ($\sigma$) of [O{\,\sc iii}]$\lambda$5007 for the five narrow components and one broad component from the best-fit model, after correcting for the instrumental broadening (Table \ref{tab: velo}). 
The result shows that the five narrow components consist of three blueshifted, one rest, and one redshifted. 
Despite their various velocity shifts, the velocity dispersions are similar. 
The broad component is blueshifted.

\section{Discussion} \label{sec:discussion}

\subsection{Comparison with Other AGN Classes} \label{subsec: comp}

%% ========
%%  figure 
%% ========
\begin{figure*}[t]
\includegraphics[width=17.7cm]{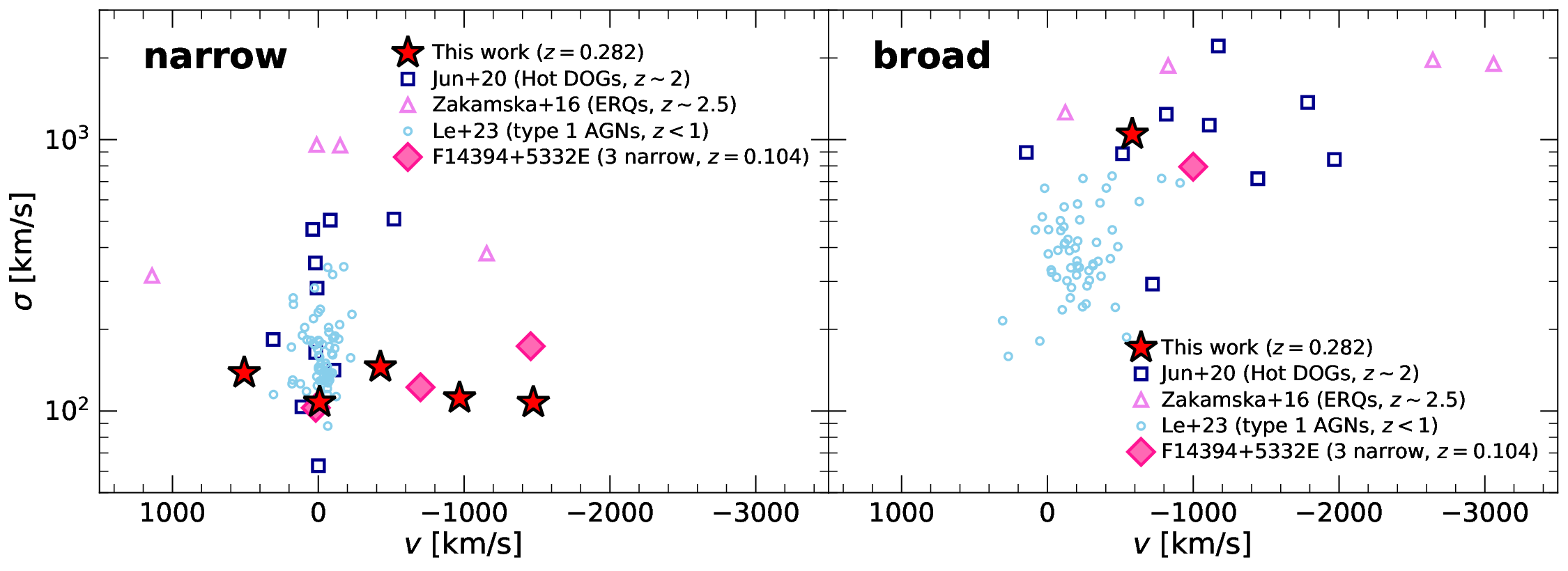}
\caption{
Velocity dispersion ($\sigma$) as a function of velocity shift ($v$) for the narrow (left) and broad (right) [O{\,\sc iii}]$\lambda$5007 components of J1010+3725 (filled red stars) and comparison samples. 
For comparison, Hot DOGs (Jun et al. \citeyear{Jun+20}; dark-blue open squares), ERQs (Zakamska et al. \citeyear{Zakamska+16}; violet open triangles), type 1 AGNs (Le et al. \citeyear{Le+23}; light-blue open circles), and IRAS F14394+5332E (Spence et al. \citeyear{Spence+18}; pink open diamonds) are plotted. 
\label{fig: v_comp}}
\end{figure*}
%% =====
%%  end
%% =====

To assess the nature of J1010+3725 within the context of other AGN populations, we compared the kinematics of its narrow and broad [O{\,\sc iii}]$\lambda$5007 components to type 1 and other dusty AGNs, based on their velocity shifts ($v$) and velocity dispersions ($\sigma$). 
To serve as comparison samples, we took Hot DOGs (DOGs with significantly red {\it WISE} colors; Eisenhardt et al. \citeyear{Eisenhardt+12}; Wu et al. \citeyear{Wu+12}) from Jun et al. (\citeyear{Jun+20}), extreme red quasars (ERQs, type 1 quasars that satisfy the DOG criterion; Ross et al. \citeyear{Ross+15}) from Zakamska et al. (\citeyear{Zakamska+16}), and type 1 AGNs from Le et al. (\citeyear{Le+23}). 
Figure \ref{fig: v_comp} (a) and (b) show the results of the comparison for the narrow and broad components, respectively. 
The five narrow components have velocity dispersions similar to those of type 1 AGNs and equal to or smaller than those of dusty AGNs. 
Interestingly, the velocity shift of the five narrow components nearly covers the full range of the comparison samples. 
The velocity dispersion of the broad component is comparable to those of dusty AGNs and larger than those of type 1 AGNs, suggesting the presence of a strong outfow in J1010+3725. 

We further compared the kinematics with those of IRAS F14394+5332E (Spence et al. \citeyear{Spence+18}), which is an ultraluminous infrared galaxy (ULIRG: $L_{\rm IR} > 10^{12} L_\odot$) having three narrow components in the [O{\,\sc iii}]$\lambda$5007 emission. 
For both the narrow and broad components, the distributions of the two objects in Figure \ref{fig: v_comp} are close, implying that their ionized-gas kinematics are comparable. 

As for the complex narrow-line structure, although it does not significantly differ from those of other AGNs in terms of velocity shift and velocity dispersion, it is clearly resolved into five distinct components. 
Although the detailed geometrical structure of those five components cannot be discussed using the currently available datasets, integral-field spectroscopic observations with a high angular resolution may reveal their spatial distribution.
As for the broad component, we are probably observing an outflowing component whose characteristics are similar to those seen in other AGNs.

\subsection{Physical Nature of the Outflowing Gas} \label{subsec: cloudy}

To investigate the physical condition of the outflowing gas, we compared the observed and model-predicted line ratios of the narrow components, specifically [O{\,\sc iii}]$\lambda$5007/[O{\,\sc iii}]$\lambda$4363 and [O{\,\sc iii}]$\lambda$5007/[Ne{\,\sc iii}]$\lambda$3868. 
The observed ratios and their uncertainties were derived from our best-fit spectral model. 
Among the five components, only the rest one yielded reliable measurements (S/N $>3$) for both line ratios.
Thus, we mainly focused on this main component to discuss the physical nature.

We performed photoionization modeling using the \texttt{CLOUDY} version 23.01 (Chatzikos et al. \citeyear{Chatzikos+23}; Gunasekera et al. \citeyear{Gunasekera+23}). 
For the calculations, we adopted gas metallicities of $Z_\odot$ and $2Z_\odot$. 
%For the calculations, we adopted a metallicity of $2Z_\odot$ since earlier works reported super-solar metallicities for NLR clouds in AGNs (e.g., Storchi-Bergmann \citeyear{Storchi+91}; Terao et al. \citeyear{Terao+22}). 
For the ionization parameter ($U$) and hydrogen density of the cloud ($n$), we set ranges of $10^{-3.5}$--$10^{-1.5}$ and $10^3$--10$^7$ cm$^{-3}$ with a step size of 1.0 dex, respectively. 
Orion-type graphite and silicate grains (Baldwin et al. \citeyear{Baldwin+91}) were included, along with the effects of dust on heavy-element depletion and radiative transfer. 
The AGN SED was assumed to be as follows:
\begin{equation}
\label{eq: photo}
f_{\nu} = \nu^{\alpha_{\rm UV}} {\rm exp}(-h\nu / kT_{\rm BB}){\rm exp}(kT_{\rm IR}/ h\nu) + b \nu^{\alpha_{\rm X}}
\end{equation}
Here, we took the two types of SED parameters originally considered for broad-line Seyfert1 galaxies (BLS1s; $\alpha_{\rm UV}=-0.5$, $T_{\rm BB}=1.18 \times 10^6$ K, $kT_{\rm IR}=0.01$ Ryd, $\alpha_{\rm X}=-1.15$, and $b$ is a constant yielding $\alpha_{\rm OX}=-1.35$) and narrow-line Seyfert1 galaxies (NLS1s; $\alpha_{\rm UV}=-0.5$, $T_{\rm BB}=4.9 \times 10^5$ K, $kT_{\rm IR}=0.01$ Ryd, $\alpha_{\rm X}=-0.85$, and $b$ is a constant yielding $\alpha_{\rm OX}=-1.35$), referring to Nagao et al. (\citeyear{Nagao+01c}).

As well as the AGN photoionization, we also considered the shock model because it can produce the low [O{\,\sc iii}]$\lambda$5007/[O{\,\sc iii}]$\lambda$4363 observed in J1010+3725 (e.g., Dopita \& Michael \citeyear{Dopita+96}; Morse et al. \citeyear{Morse+96}; Leung et al. \citeyear{Leung+21}). 
However, the shock models in wide parameter ranges studied by Dopita \& Sutherland (\citeyear{Dopita+96}) predict a moderately high flux ratio of ([O{\,\sc ii}]$\lambda$3727+[O{\,\sc ii}]$\lambda$3729)/[O{\,\sc iii}]$\lambda$5007 ($>3$). 
Nevertheless, our target shows no multiple peaks in [O{\,\sc ii}]$\lambda$$\lambda$3727, 3729 doublet emission, suggesting that the multi-components of outflow do not originate from the shock. 
AGN-photoionized dense gas clouds are associated with a significantly low flux ratio of ([O{\,\sc ii}]$\lambda$3727+[O{\,\sc ii}]$\lambda$3729)/[O{\,\sc iii}]$\lambda$5007 because the relative emissivity of the [O{\,\sc ii}] lines decreases in high-density gas compared with that of [O{\,\sc iii}]$\lambda$5008 (i.e., $n > 10^4$ cm$^{-3}$). 
Also, highly AGN-photoionized gas clouds can account for the significantly low flux ratio of ([O{\,\sc ii}]$\lambda$3727+[O{\,\sc ii}]$\lambda$3729)/[O{\,\sc iii}]$\lambda$5007 in J1010+3725. 
Therefore, in our model calculations, we assumed the AGN-photoionized gas with a wide range of ionization parameters ($U$) and hydrogen densities of the cloud ($n$). 
All of our calculations were stopped at temperature of 4000 K (i.e., assuming the radiation-bound model). 
It has been reported that the matter-bound model can produce low [O{\,\sc iii}]$\lambda$5007/[O{\,\sc iii}]$\lambda$4363 flux ratios. 
However, the minimum value is $\sim$ 40  for the flux ratio of [O{\,\sc iii}]$\lambda$5007/[O{\,\sc iii}]$\lambda$4363 (Binette et al. \citeyear{Binette+96}), which is relatively higher than the extinction-corrected observed values. 
Furthermore, the matter-bound model produces relatively strong high-ionization lines (e.g., He{\,\sc ii}$\lambda$4686 and [Ne{\,\sc v}]$\lambda$3425) compared with the radiation-bound model. 
However, such lines are not detected in J1010+3725 (see Figure \ref{fig: spec}). 
Therefore, we assumed radiation-bound model.

Dust extinction was estimated from the observed Balmer decrement. 
The measured narrow-line H$\gamma$/H$\beta$ flux ratio, $0.337^{+0.106}_{-0.116}$, was lower than the expected value of H$\gamma$/H$\beta$ = 0.469 for case B recombination at $T$ = 10$^4$ K and $n$ = 10$^4$ cm$^{-3}$ (Hummer \& Storey \citeyear{Hummer+87}). 
The best-fit value means substantial extinction on the observed line ratios, but with large uncertainties due to the significant blending in the narrow-line H$\gamma$ (see Section \ref{subsec: fit}); thus, the possibility of little extinction cannot be ruled out.
Using the extinction curve of Cardelli et al. (\citeyear{Cardelli+89}) with $R_V = 3.1$, we estimated a dust extinction of $A_V =1.97^{+1.42}_{-1.57}$ mag, which is consistent with the typical value of $A_V=2.1$ for SDSS DOGs reported by Toba et al. (\citeyear{Toba+17}). 
%% ==========
%%  footnote 
%% ==========
\footnote{Toba et al. (\citeyear{Toba+17}) performed SED fitting for DOGs using SDSS and {\it WISE} photometries and obtained a value of $E(B-V) = 0.68$. 
Assuming $R_V = 3.1$, this value is converted to $A_V = 2.1$ mag.}
%% =====
%%  end
%% =====
Although the H$\alpha$/H$\beta$ flux ratio is often used to estimate $A_V$, the strong broad H$\alpha$ emission makes it difficult to decompose the narrow H$\alpha$ component from the broad component in J1010+3725  (see Figure \ref{fig: spec} (e)). 
Therefore, we adopted H$\gamma$/H$\beta$ to infer $A_V$.

%% ========
%%  figure 
%% ========
\begin{figure*}[t]
\includegraphics[width=17.7cm]{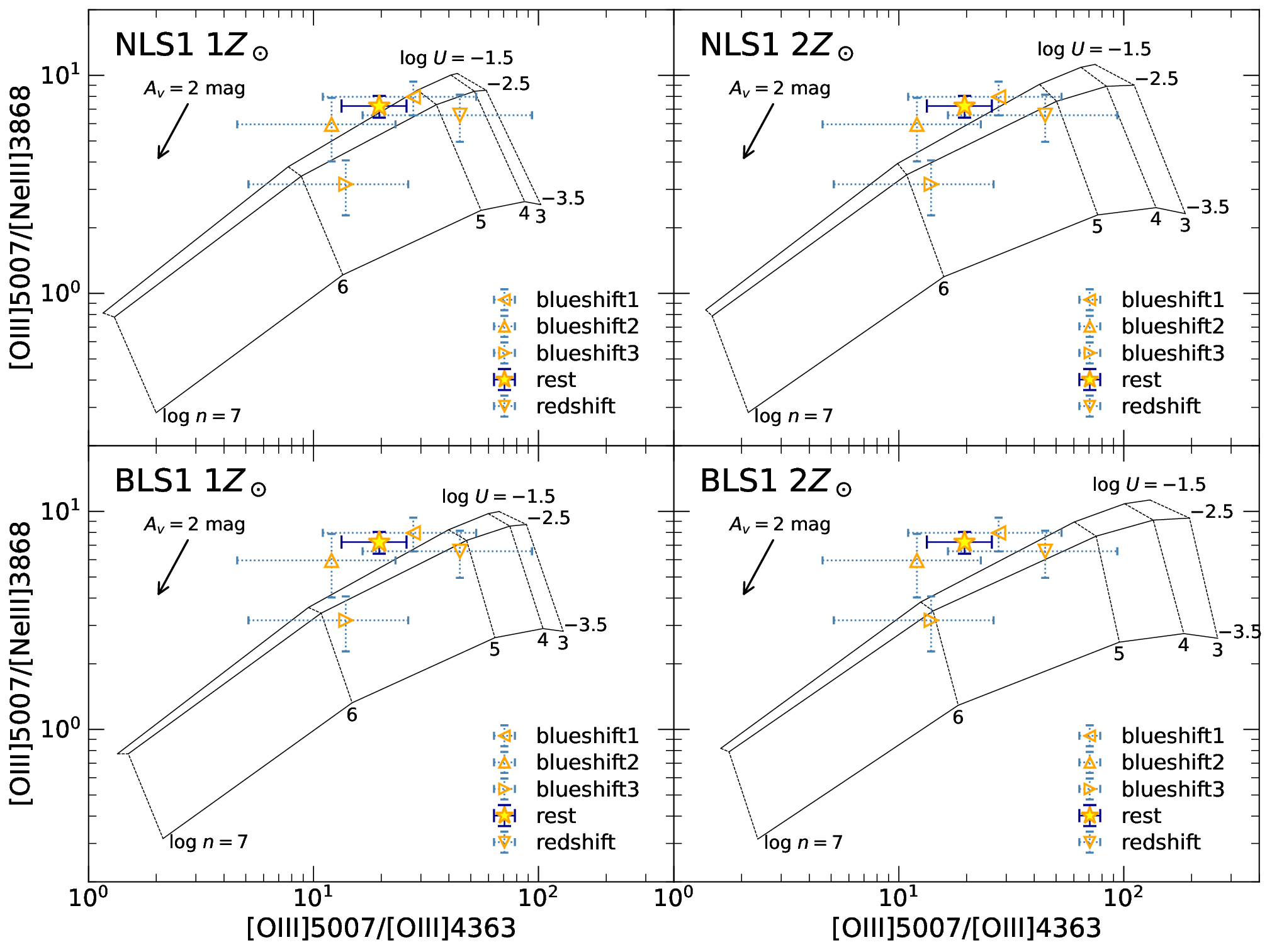}
\caption{
Comparison between the non extinction-corrected) observed and modeled line ratios for [O{\,\sc iii}]$\lambda$5007/[O{\,\sc iii}]$\lambda$4363 and [O{\,\sc iii}]$\lambda$5007/[Ne{\,\sc iii}]$\lambda$3868. 
The upper and lower panels represent the NLS1 and BLS1 SED models, respectively. 
The models with $Z_\odot$ and $2Z_\odot$ are shown in the left and right panels, respectively. 
The star and triangles represent the observed flux ratios with reliable (S/N $>3$) and unreliable measurements, respectively. 
The blueshifted components with the highest, intermediate, and lowest velocity shifts are referred to as blueshift1, blueshift2, and blueshift3 in the legend, respectively. 
The solid and dotted lines depict photoionization model grids with constant $U$ and $n$, respectively. 
The black arrow indicates the correction of dust extinction with $A_V = 2$. 
\label{fig: line_ratio}}
\end{figure*}
%% =====
%%  end
%% =====

Figure \ref{fig: line_ratio} shows a comparison between the observed and modeled emission-line flux ratios. 
The observed line ratios lie above or on the edge of the model grid. 
Assuming the best-fit value of dust extinction ($A_V = 1.97$), these results indicate that the rest component is characterized by a high density of $\sim 10^{6.0}$ cm$^{-3}$ and a relatively high ionization parameter of $\sim 10^{-2.5}$, regardless of the assumed SED and metallicity. 
Even when accounting for the large uncertainty in dust extinction ($A_V =1.97^{+1.42}_{-1.57}$), a high density of $\sim$10$^{5.5}$--10$^{6.0}$ cm$^{-3}$ is still required for the rest component, whereas the ionization parameter can span a wide range of $\sim 10^{-3.0}$--$10^{-1.5}$. 
For the other components, the best-fit dust extinction results in similarly high densities ($\sim$10$^{5.5}$--10$^{6.5}$ cm$^{-3}$), but with larger uncertainties. 
To discuss the physical property of the blueshift and redshift components in more detail, deeper spectroscopic observations are required.

Recent investigations of gas densities using some diagnostic emission-line flux ratios (e.g., [S{\,\sc ii}]$\lambda$4068/[S{\,\sc ii}]$\lambda$4076, [S{\,\sc ii}]$\lambda$6716/[S{\,\sc ii}]$\lambda$6731, [O{\,\sc ii}]$\lambda$7319/[O{\,\sc ii}]$\lambda$7331, and [Ne{\,\sc V}]14.3 $\mu$m/[Ne{\,\sc V}]24.3 $\mu$m) evidence high outflow densities in AGNs ($\sim$ $10^3$--$10^5$ cm$^{-3}$; e.g., Rose et al. \citeyear{Rose+18}, Spence et al. \citeyear{Spence+18}, Romas Almeida et al. \citeyear{Ramos+25}), but not as high as our estimation.
Therefore, the five outflow components in J1010+3725 represent an extreme case of not only complex kinematics but also high gas density. 

Our results imply that the five highly dense gas components are outflowing with multiple (line-of-sight) bulk velocities. 
On the narrow-line region (NLR) scale, outflows are mostly driven by radiation pressure from AGNs, and theoretical models in which they are launched from the dusty torus as a dusty wind have also been proposed (see, e.g., Kudoh et al. \citeyear{Kudoh+23}). 
The dense gas observed in J1010+3725 tends to be located in the innermost part of the NLR (Nagao et al. \citeyear{Nagao+01a}, \citeyear{Nagao+01b}), meaning that the outflow of dense dusty gas observed in J1010+3725 is probably in a very early stage after the launch near the torus (e.g., Dorodnitsyn et al. \citeyear{Dorodnitsyn+16}). 
This suggests that the observed multiple peaks of the outflow may corresponds to the inhomogeneous or clumpy structure of the torus (e.g., Nenkova et al. \citeyear{Nenkova+08a}, \citeyear{Nenkova+08b}; Wada \citeyear{Wada+12}, \citeyear{Wada+16}; Tanimoto et al. \citeyear{Tanimoto+19}).

Tadhunter et al. (\citeyear{Tadhunter+18}, \citeyear{Tadhunter+19}) analyzed the {\it Hubble Space Telescope} ({\it HST}) narrow-band [O{\,\sc iii}]$\lambda$5007 imaging and integral-field spectroscopy (IFS) obtained from {\it HST}/{\it Space Telescope Imaging Spectrograph} ({\it STIS}) for IRAS F14394+5332E (with three peaks in [O{\,\sc iii}]$\lambda$5007). 
They reported an extended [O{\,\sc iii}]$\lambda$5007-emitting region ($\sim$ 0.9 kpc) and the absence of a concentrated [O{\,\sc iii}]$\lambda$5007 emission around the AGN. 
These results are in contrast to our scenario for the five narrow components in J1010+3725, implying a wide variation in the origin of the multi-peak [O{\,\sc iii}]$\lambda$5007 emission. 
To test our results for the five narrow components in J1010+3725 and to clarify the difference between J1010+3725 and F14394+5332E, detailed IFS observations of J1010+3725 are required.  
Furthermore, a statistically large sample of multi-peak [O{\,\sc iii}]$\lambda$5007 emitters is required to understand the wide variation in the nature of the AGN outflow in the spatial scale of NLRs.

%\section{Conclusion} \label{sec:conlusion}

%TBD

%\begin{enumerate}
%\item AAA
%\end{enumerate}

\section*{acknowledgments}
%\begin{acknowledgments}

The authors gratefully acknowledge the anonymous referee for a careful reading of the manuscript and very helpful comments.

%SDSS
Funding for the Sloan Digital Sky Survey V has been provided by the Alfred P. Sloan Foundation, the Heising-Simons Foundation, the National Science Foundation, and the Participating Institutions. SDSS acknowledges support and resources from the Center for High-Performance Computing at the University of Utah. SDSS telescopes are located at Apache Point Observatory, funded by the Astrophysical Research Consortium and operated by New Mexico State University, and at Las Campanas Observatory, operated by the Carnegie Institution for Science. The SDSS web site is \url{www.sdss.org}.
SDSS is managed by the Astrophysical Research Consortium for the Participating Institutions of the SDSS Collaboration, including the Carnegie Institution for Science, Chilean National Time Allocation Committee (CNTAC) ratified researchers, Caltech, the Gotham Participation Group, Harvard University, Heidelberg University, The Flatiron Institute, The Johns Hopkins University, L'Ecole polytechnique f\'{e}d\'{e}rale de Lausanne (EPFL), Leibniz-Institut f\"{u}r Astrophysik Potsdam (AIP), Max-Planck-Institut f\"{u}r Astronomie (MPIA Heidelberg), Max-Planck-Institut f\"{u}r Extraterrestrische Physik (MPE), Nanjing University, National Astronomical Observatories of China (NAOC), New Mexico State University, The Ohio State University, Pennsylvania State University, Smithsonian Astrophysical Observatory, Space Telescope Science Institute (STScI), the Stellar Astrophysics Participation Group, Universidad Nacional Aut\'{o}noma de M\'{e}xico, University of Arizona, University of Colorado Boulder, University of Illinois at Urbana-Champaign, University of Toronto, University of Utah, University of Virginia, Yale University, and Yunnan University.

%GALEX
Based on observations made with the NASA Galaxy Evolution Explorer.
$GALEX$ is operated for NASA by the California Institute of Technology under NASA contract NAS5-98034.

%2MASS
This publication makes use of data products from the Two Micron All Sky Survey, which is a joint project of the University of Massachusetts and the Infrared Processing and Analysis Center/California Institute of Technology, funded by the National Aeronautics and Space Administration and the National Science Foundation.

%WISE
This publication makes use of data products from the Wide-field Infrared Survey Explorer, which is a joint project of the University of California, Los Angeles, and the Jet Propulsion Laboratory/California Institute of Technology, funded by the National Aeronautics and Space Administration.

%AKARI
This research is based on observations with {\it AKARI}, a JAXA project with the participation of ESA.

This work was financially supported by JSPS KAKENHI, grant Nos. 20H01949 (T.N.), 23H01215 (T.N.), 23K22537 (Y.T.) and 24KJ1833 (N.Y.), and by JST SPRING, grant No. JPMJSP2162 (K.S.). 

%\end{acknowledgments}

%appendix

\bibliography{OIII_DOG}{}

@ARTICLE{Baldwin+91,
       author = {{Baldwin}, Jack A. and {Ferland}, Gary J. and {Martin}, P.~G. and {Corbin}, Michael R. and {Cota}, Stephen A. and {Peterson}, Bradley M. and {Slettebak}, Arne},
        title = "{Physical Conditions in the Orion Nebula and an Assessment of Its Helium Abundance}",
      journal = {\apj},
         year = 1991,
        month = jun,
       volume = {374},
        pages = {580},
          doi = {10.1086/170146},
       adsurl = {https://ui.adsabs.harvard.edu/abs/1991ApJ...374..580B}
}

@ARTICLE{Barnes+91,
       author = {{Barnes}, Joshua E. and {Hernquist}, Lars E.},
        title = "{Fueling Starburst Galaxies with Gas-rich Mergers}",
      journal = {\apjl},
         year = 1991,
        month = apr,
       volume = {370},
        pages = {L65},
          doi = {10.1086/185978},
       adsurl = {https://ui.adsabs.harvard.edu/abs/1991ApJ...370L..65B}
}

@ARTICLE{Becker+95,
       author = {{Becker}, Robert H. and {White}, Richard L. and {Helfand}, David J.},
        title = "{The FIRST Survey: Faint Images of the Radio Sky at Twenty Centimeters}",
      journal = {\apj},
         year = 1995,
        month = sep,
       volume = {450},
        pages = {559},
          doi = {10.1086/176166},
       adsurl = {https://ui.adsabs.harvard.edu/abs/1995ApJ...450..559B}
}

@ARTICLE{Binette+96,
       author = {{Binette}, L. and {Wilson}, A.~S. and {Storchi-Bergmann}, T.},
        title = "{Excitation and temperature of extended gas in active galaxies. II. Photoionization models with matter-bounded clouds.}",
      journal = {\aap},
         year = 1996,
        month = aug,
       volume = {312},
        pages = {365-379},
       adsurl = {https://ui.adsabs.harvard.edu/abs/1996A&A...312..365B}
}

@ARTICLE{Bussmann+09,
       author = {{Bussmann}, R.~S. and {Dey}, Arjun and {Borys}, C. and {Desai}, V. and {Jannuzi}, B.~T. and {Le Floc'h}, E. and {Melbourne}, J. and {Sheth}, K. and {Soifer}, B.~T.},
        title = "{Infrared Luminosities and Dust Properties of z {\ensuremath{\approx}} 2 Dust-obscured Galaxies}",
      journal = {\apj},
         year = 2009,
        month = nov,
       volume = {705},
       number = {1},
        pages = {184-198},
          doi = {10.1088/0004-637X/705/1/184},
archivePrefix = {arXiv},
       eprint = {0909.2650},
 primaryClass = {astro-ph.CO},
       adsurl = {https://ui.adsabs.harvard.edu/abs/2009ApJ...705..184B}
}

@ARTICLE{Cardelli+89,
       author = {{Cardelli}, Jason A. and {Clayton}, Geoffrey C. and {Mathis}, John S.},
        title = "{The Relationship between Infrared, Optical, and Ultraviolet Extinction}",
      journal = {\apj},
         year = 1989,
        month = oct,
       volume = {345},
        pages = {245},
          doi = {10.1086/167900},
       adsurl = {https://ui.adsabs.harvard.edu/abs/1989ApJ...345..245C}
}

@ARTICLE{Chatzikos+23,
       author = {{Chatzikos}, M. and {Bianchi}, S. and {Camilloni}, F. and {Chakraborty}, P. and {Gunasekera}, C.~M. and {Guzm{\'a}n}, F. and {Milby}, J.~S. and {Sarkar}, A. and {Shaw}, G. and {van Hoof}, P.~A.~M. and {Ferland}, G.~J.},
        title = "{The 2023 Release of Cloudy}",
      journal = {\rmxaa},
         year = 2023,
        month = oct,
       volume = {59},
        pages = {327-343},
          doi = {10.22201/ia.01851101p.2023.59.02.12},
archivePrefix = {arXiv},
       eprint = {2308.06396},
 primaryClass = {astro-ph.GA},
       adsurl = {https://ui.adsabs.harvard.edu/abs/2023RMxAA..59..327C}
}

@dataset{Cutri+21,
       author = {{Cutri}, R.~M. and {Wright}, E.~L. and {Conrow}, T. and {Fowler}, J.~W. and {Eisenhardt}, P.~R.~M. and {Grillmair}, C. and {Kirkpatrick}, J.~D. and {Masci}, F. and {McCallon}, H.~L. and {Wheelock}, S.~L. and {Fajardo-Acosta}, S. and {Yan}, L. and {Benford}, D. and {Harbut}, M. and {Jarrett}, T. and {Lake}, S. and {Leisawitz}, D. and {Ressler}, M.~E. and {Stanford}, S.~A. and {Tsai}, C. -W. and {Liu}, F. and {Helou}, G. and {Mainzer}, A. and {Gettngs}, D. and {Gonzalez}, A. and {Hoffman}, D. and {Marsh}, K.~A. and {Padgett}, D. and {Skrutskie}, M.~F. and {Beck}, R. and {Papin}, M. and {Wittman}, M.},
        title = "{VizieR Online Data Catalog: AllWISE Data Release (Cutri+ 2013)}",
 howpublished = {VizieR On-line Data Catalog: II/328.  Originally published in: IPAC/Caltech (2013)},
         year = 2021,
        month = feb,
          eid = {II/328},
       adsurl = {https://ui.adsabs.harvard.edu/abs/2014yCat.2328....0C}
}

@ARTICLE{Desai+09,
       author = {{Desai}, Vandana and {Soifer}, B.~T. and {Dey}, Arjun and {Le Floc'h}, Emeric and {Armus}, Lee and {Brand}, Kate and {Brown}, Michael J.~I. and {Brodwin}, Mark and {Jannuzi}, Buell T. and {Houck}, James R. and {Weedman}, Daniel W. and {Ashby}, Matthew L.~N. and {Gonzalez}, Anthony and {Huang}, Jiasheng and {Smith}, Howard A. and {Teplitz}, Harry and {Willner}, Steve P. and {Melbourne}, Jason},
        title = "{Strong Polycyclic Aromatic Hydrocarbon Emission from z {\ensuremath{\approx}} 2 ULIRGs}",
      journal = {\apj},
         year = 2009,
        month = aug,
       volume = {700},
       number = {2},
        pages = {1190-1204},
          doi = {10.1088/0004-637X/700/2/1190},
archivePrefix = {arXiv},
       eprint = {0905.4274},
 primaryClass = {astro-ph.CO},
       adsurl = {https://ui.adsabs.harvard.edu/abs/2009ApJ...700.1190D}
}

@ARTICLE{Dey+08,
       author = {{Dey}, Arjun and {Soifer}, B.~T. and {Desai}, Vandana and {Brand}, Kate and {Le Floc'h}, Emeric and {Brown}, Michael J.~I. and {Jannuzi}, Buell T. and {Armus}, Lee and {Bussmann}, Shane and {Brodwin}, Mark and {Bian}, Chao and {Eisenhardt}, Peter and {Higdon}, Sarah J. and {Weedman}, Daniel and {Willner}, S.~P.},
        title = "{A Significant Population of Very Luminous Dust-Obscured Galaxies at Redshift z \raisebox{-0.5ex}\textasciitilde 2}",
      journal = {\apj},
         year = 2008,
        month = apr,
       volume = {677},
       number = {2},
        pages = {943-956},
          doi = {10.1086/529516},
archivePrefix = {arXiv},
       eprint = {0801.1860},
 primaryClass = {astro-ph},
       adsurl = {https://ui.adsabs.harvard.edu/abs/2008ApJ...677..943D}
}

@ARTICLE{Dopita+96,
       author = {{Dopita}, Michael A. and {Sutherland}, Ralph S.},
        title = "{Spectral Signatures of Fast Shocks. I. Low-Density Model Grid}",
      journal = {\apjs},
         year = 1996,
        month = jan,
       volume = {102},
        pages = {161},
          doi = {10.1086/192255},
       adsurl = {https://ui.adsabs.harvard.edu/abs/1996ApJS..102..161D}
}

@ARTICLE{Dorodnitsyn+16,
       author = {{Dorodnitsyn}, A. and {Kallman}, T. and {Proga}, D.},
        title = "{Parsec-scale Accretion and Winds Irradiated by a Quasar}",
      journal = {\apj},
         year = 2016,
        month = mar,
       volume = {819},
       number = {2},
          eid = {115},
        pages = {115},
          doi = {10.3847/0004-637X/819/2/115},
archivePrefix = {arXiv},
       eprint = {1512.03854},
 primaryClass = {astro-ph.GA},
       adsurl = {https://ui.adsabs.harvard.edu/abs/2016ApJ...819..115D}
}

@ARTICLE{Eisenhardt+12,
       author = {{Eisenhardt}, Peter R.~M. and {Wu}, Jingwen and {Tsai}, Chao-Wei and {Assef}, Roberto and {Benford}, Dominic and {Blain}, Andrew and {Bridge}, Carrie and {Condon}, J.~J. and {Cushing}, Michael C. and {Cutri}, Roc and {Evans}, II, Neal J. and {Gelino}, Chris and {Griffith}, Roger L. and {Grillmair}, Carl J. and {Jarrett}, Tom and {Lonsdale}, Carol J. and {Masci}, Frank J. and {Mason}, Brian S. and {Petty}, Sara and {Sayers}, Jack and {Stanford}, S.~A. and {Stern}, Daniel and {Wright}, Edward L. and {Yan}, Lin},
        title = "{The First Hyper-luminous Infrared Galaxy Discovered by WISE}",
      journal = {\apj},
         year = 2012,
        month = aug,
       volume = {755},
       number = {2},
          eid = {173},
        pages = {173},
          doi = {10.1088/0004-637X/755/2/173},
archivePrefix = {arXiv},
       eprint = {1208.5517},
 primaryClass = {astro-ph.CO},
       adsurl = {https://ui.adsabs.harvard.edu/abs/2012ApJ...755..173E}
}

@ARTICLE{Fan+17,
       author = {{Fan}, Lulu and {Jones}, Suzy F. and {Han}, Yunkun and {Knudsen}, Kirsten K.},
        title = "{The SCUBA-2 850 {\ensuremath{\mu}}m Follow-up of WISE-selected, Luminous Dust-obscured Quasars}",
      journal = {\pasp},
         year = 2017,
        month = dec,
       volume = {129},
       number = {982},
        pages = {124101},
          doi = {10.1088/1538-3873/aa8e91},
archivePrefix = {arXiv},
       eprint = {1709.07971},
 primaryClass = {astro-ph.GA},
       adsurl = {https://ui.adsabs.harvard.edu/abs/2017PASP..129l4101F}
}

@ARTICLE{Finnerty+20,
       author = {{Finnerty}, Luke and {Larson}, Kirsten and {Soifer}, B.~T. and {Armus}, Lee and {Matthews}, Keith and {Jun}, Hyunsung D. and {Moon}, Dae-Sik and {Melbourne}, Jason and {Gomez}, Percy and {Tsai}, Chao-Wei and {D{\'\i}az-Santos}, Tanio and {Eisenhardt}, Peter and {Cushing}, Michael},
        title = "{Fast Outflows in Hot Dust-obscured Galaxies Detected with Keck/NIRES}",
      journal = {\apj},
         year = 2020,
        month = dec,
       volume = {905},
       number = {1},
          eid = {16},
        pages = {16},
          doi = {10.3847/1538-4357/abc3bf},
archivePrefix = {arXiv},
       eprint = {2010.10641},
 primaryClass = {astro-ph.GA},
       adsurl = {https://ui.adsabs.harvard.edu/abs/2020ApJ...905...16F}
}

@ARTICLE{Foreman-Mackey+13,
       author = {{Foreman-Mackey}, Daniel and {Hogg}, David W. and {Lang}, Dustin and {Goodman}, Jonathan},
        title = "{emcee: The MCMC Hammer}",
      journal = {\pasp},
         year = 2013,
        month = mar,
       volume = {125},
       number = {925},
        pages = {306},
          doi = {10.1086/670067},
archivePrefix = {arXiv},
       eprint = {1202.3665},
 primaryClass = {astro-ph.IM},
       adsurl = {https://ui.adsabs.harvard.edu/abs/2013PASP..125..306F}
}

@ARTICLE{Gunasekera+23,
       author = {{Gunasekera}, Chamani M. and {van Hoof}, Peter A.~M. and {Chatzikos}, Marios and {Ferland}, Gary J.},
        title = "{The 23.01 Release of Cloudy}",
      journal = {Research Notes of the American Astronomical Society},
         year = 2023,
        month = nov,
       volume = {7},
       number = {11},
          eid = {246},
        pages = {246},
          doi = {10.3847/2515-5172/ad0e75},
archivePrefix = {arXiv},
       eprint = {2311.10163},
 primaryClass = {astro-ph.GA},
       adsurl = {https://ui.adsabs.harvard.edu/abs/2023RNAAS...7..246G}
}

@ARTICLE{Hopkins+08,
       author = {{Hopkins}, Philip F. and {Hernquist}, Lars and {Cox}, Thomas J. and {Kere{\v{s}}}, Du{\v{s}}an},
        title = "{A Cosmological Framework for the Co-Evolution of Quasars, Supermassive Black Holes, and Elliptical Galaxies. I. Galaxy Mergers and Quasar Activity}",
      journal = {\apjs},
         year = 2008,
        month = apr,
       volume = {175},
       number = {2},
        pages = {356-389},
          doi = {10.1086/524362},
archivePrefix = {arXiv},
       eprint = {0706.1243},
 primaryClass = {astro-ph},
       adsurl = {https://ui.adsabs.harvard.edu/abs/2008ApJS..175..356H}
}

@ARTICLE{Hummer+87,
       author = {{Hummer}, D.~G. and {Storey}, P.~J.},
        title = "{Recombination-line intensities for hydrogenic ions - I. Case B calculations for H I and He II.}",
      journal = {\mnras},
         year = 1987,
        month = feb,
       volume = {224},
        pages = {801-820},
          doi = {10.1093/mnras/224.3.801},
       adsurl = {https://ui.adsabs.harvard.edu/abs/1987MNRAS.224..801H}
}

@ARTICLE{Jun+20,
       author = {{Jun}, Hyunsung D. and {Assef}, Roberto J. and {Bauer}, Franz E. and {Blain}, Andrew and {D{\'\i}az-Santos}, Tanio and {Eisenhardt}, Peter R.~M. and {Stern}, Daniel and {Tsai}, Chao-Wei and {Wright}, L., Edward and {Wu}, Jingwen},
        title = "{Spectral Classification and Ionized Gas Outflows in z {\ensuremath{\sim}} 2 WISE-selected Hot Dust-obscured Galaxies}",
      journal = {\apj},
         year = 2020,
        month = jan,
       volume = {888},
       number = {2},
          eid = {110},
        pages = {110},
          doi = {10.3847/1538-4357/ab5e7b},
archivePrefix = {arXiv},
       eprint = {1911.09828},
 primaryClass = {astro-ph.GA},
       adsurl = {https://ui.adsabs.harvard.edu/abs/2020ApJ...888..110J}
}

@ARTICLE{Kudoh+23,
       author = {{Kudoh}, Yuki and {Wada}, Keiichi and {Kawakatu}, Nozomu and {Nomura}, Mariko},
        title = "{Multiphase Gas Nature in the Sub-parsec Region of the Active Galactic Nuclei. I. Dynamical Structures of Dusty and Dust-free Outflow}",
      journal = {\apj},
         year = 2023,
        month = jun,
       volume = {950},
       number = {1},
          eid = {72},
        pages = {72},
          doi = {10.3847/1538-4357/accc2b},
archivePrefix = {arXiv},
       eprint = {2304.05950},
 primaryClass = {astro-ph.GA},
       adsurl = {https://ui.adsabs.harvard.edu/abs/2023ApJ...950...72K}
}

@ARTICLE{Kormendy+13,
       author = {{Kormendy}, John and {Ho}, Luis C.},
        title = "{Coevolution (Or Not) of Supermassive Black Holes and Host Galaxies}",
      journal = {\araa},
         year = 2013,
        month = aug,
       volume = {51},
       number = {1},
        pages = {511-653},
          doi = {10.1146/annurev-astro-082708-101811},
archivePrefix = {arXiv},
       eprint = {1304.7762},
 primaryClass = {astro-ph.CO},
       adsurl = {https://ui.adsabs.harvard.edu/abs/2013ARA&A..51..511K}
}

@ARTICLE{Le+23,
       author = {{Le}, Huynh Anh N. and {Xue}, Yongquan and {Lin}, Xiaozhi and {Wang}, Yijun},
        title = "{[O III] 5007 {\r{A}} Emission Line Width as a Surrogate for {\ensuremath{\sigma}} $_{{\ensuremath{*}}}$ in Type 1 AGNs?}",
      journal = {\apj},
         year = 2023,
        month = mar,
       volume = {945},
       number = {1},
          eid = {59},
        pages = {59},
          doi = {10.3847/1538-4357/acb770},
archivePrefix = {arXiv},
       eprint = {2301.12918},
 primaryClass = {astro-ph.GA},
       adsurl = {https://ui.adsabs.harvard.edu/abs/2023ApJ...945...59L}
}

@ARTICLE{Leung+21,
       author = {{Leung}, Gene C.~K. and {Coil}, Alison L. and {Rupke}, David S.~N. and {Perrotta}, Serena},
        title = "{KCWI Observations of the Extended Nebulae in Mrk 273}",
      journal = {\apj},
         year = 2021,
        month = jun,
       volume = {914},
       number = {1},
          eid = {17},
        pages = {17},
          doi = {10.3847/1538-4357/abf4da},
archivePrefix = {arXiv},
       eprint = {2011.09587},
 primaryClass = {astro-ph.GA},
       adsurl = {https://ui.adsabs.harvard.edu/abs/2021ApJ...914...17L}
}

@ARTICLE{Li+24,
       author = {{Li}, Guodong and {Assef}, Roberto J. and {Tsai}, Chao-Wei and {Wu}, Jingwen and {Eisenhardt}, Peter R.~M. and {Stern}, Daniel and {D{\'\i}az-Santos}, Tanio and {Blain}, Andrew W. and {Jun}, Hyunsung D. and {Fern{\'a}ndez Aranda}, Rom{\'a}n and {Zewdie}, Dejene},
        title = "{Black Hole Mass and Eddington Ratio Distribution of Hot Dust-obscured Galaxies}",
      journal = {\apj},
         year = 2024,
        month = aug,
       volume = {971},
       number = {1},
          eid = {40},
        pages = {40},
          doi = {10.3847/1538-4357/ad5317},
archivePrefix = {arXiv},
       eprint = {2405.20479},
 primaryClass = {astro-ph.GA},
       adsurl = {https://ui.adsabs.harvard.edu/abs/2024ApJ...971...40L}
}

@ARTICLE{Liu+10,
       author = {{Liu}, Xin and {Shen}, Yue and {Strauss}, Michael A. and {Greene}, Jenny E.},
        title = "{Type 2 Active Galactic Nuclei with Double-Peaked [O III] Lines: Narrow-Line Region Kinematics or Merging Supermassive Black Hole Pairs?}",
      journal = {\apj},
         year = 2010,
        month = jan,
       volume = {708},
       number = {1},
        pages = {427-434},
          doi = {10.1088/0004-637X/708/1/427},
archivePrefix = {arXiv},
       eprint = {0908.2426},
 primaryClass = {astro-ph.CO},
       adsurl = {https://ui.adsabs.harvard.edu/abs/2010ApJ...708..427L}
}

@ARTICLE{Magorrian+98,
       author = {{Magorrian}, John and {Tremaine}, Scott and {Richstone}, Douglas and {Bender}, Ralf and {Bower}, Gary and {Dressler}, Alan and {Faber}, S.~M. and {Gebhardt}, Karl and {Green}, Richard and {Grillmair}, Carl and {Kormendy}, John and {Lauer}, Tod},
        title = "{The Demography of Massive Dark Objects in Galaxy Centers}",
      journal = {\aj},
         year = 1998,
        month = jun,
       volume = {115},
       number = {6},
        pages = {2285-2305},
          doi = {10.1086/300353},
archivePrefix = {arXiv},
       eprint = {astro-ph/9708072},
 primaryClass = {astro-ph},
       adsurl = {https://ui.adsabs.harvard.edu/abs/1998AJ....115.2285M}
}

@ARTICLE{Marconi+03,
       author = {{Marconi}, Alessandro and {Hunt}, Leslie K.},
        title = "{The Relation between Black Hole Mass, Bulge Mass, and Near-Infrared Luminosity}",
      journal = {\apjl},
         year = 2003,
        month = may,
       volume = {589},
       number = {1},
        pages = {L21-L24},
          doi = {10.1086/375804},
archivePrefix = {arXiv},
       eprint = {astro-ph/0304274},
 primaryClass = {astro-ph},
       adsurl = {https://ui.adsabs.harvard.edu/abs/2003ApJ...589L..21M}
}

@ARTICLE{Martin+05,
       author = {{Martin}, D. Christopher and {Fanson}, James and {Schiminovich}, David and {Morrissey}, Patrick and {Friedman}, Peter G. and {Barlow}, Tom A. and {Conrow}, Tim and {Grange}, Robert and {Jelinsky}, Patrick N. and {Milliard}, Bruno and {Siegmund}, Oswald H.~W. and {Bianchi}, Luciana and {Byun}, Yong-Ik and {Donas}, Jose and {Forster}, Karl and {Heckman}, Timothy M. and {Lee}, Young-Wook and {Madore}, Barry F. and {Malina}, Roger F. and {Neff}, Susan G. and {Rich}, R. Michael and {Small}, Todd and {Surber}, Frank and {Szalay}, Alex S. and {Welsh}, Barry and {Wyder}, Ted K.},
        title = "{The Galaxy Evolution Explorer: A Space Ultraviolet Survey Mission}",
      journal = {\apjl},
         year = 2005,
        month = jan,
       volume = {619},
       number = {1},
        pages = {L1-L6},
          doi = {10.1086/426387},
archivePrefix = {arXiv},
       eprint = {astro-ph/0411302},
 primaryClass = {astro-ph},
       adsurl = {https://ui.adsabs.harvard.edu/abs/2005ApJ...619L...1M}
}

@ARTICLE{Melbourne+12,
       author = {{Melbourne}, J. and {Soifer}, B.~T. and {Desai}, Vandana and {Pope}, Alexandra and {Armus}, Lee and {Dey}, Arjun and {Bussmann}, R.~S. and {Jannuzi}, B.~T. and {Alberts}, Stacey},
        title = "{The Spectral Energy Distributions and Infrared Luminosities of z {\ensuremath{\approx}} 2 Dust-obscured Galaxies from Herschel and Spitzer}",
      journal = {\aj},
         year = 2012,
        month = may,
       volume = {143},
       number = {5},
          eid = {125},
        pages = {125},
          doi = {10.1088/0004-6256/143/5/125},
archivePrefix = {arXiv},
       eprint = {1203.3199},
 primaryClass = {astro-ph.CO},
       adsurl = {https://ui.adsabs.harvard.edu/abs/2012AJ....143..125M}
}

@ARTICLE{Morse+96,
       author = {{Morse}, Jon A. and {Raymond}, John C. and {Wilson}, Andrew S.},
        title = "{On the Viability of Gaseous Ioniziation in Active Galaxies by Fast Shocks}",
      journal = {\pasp},
         year = 1996,
        month = may,
       volume = {108},
        pages = {426},
          doi = {10.1086/133744},
       adsurl = {https://ui.adsabs.harvard.edu/abs/1996PASP..108..426M}
}

@BOOK{Moshir+92,
       author = {{Moshir}, M. and {Kopman}, G. and {Conrow}, T.~A.~O.},
        title = "{IRAS Faint Source Survey, Explanatory supplement version 2}",
         year = 1992,
       adsurl = {https://ui.adsabs.harvard.edu/abs/1992ifss.book.....M}
}

@ARTICLE{Murakami+07,
       author = {{Murakami}, Hiroshi and {Baba}, Hajime and {Barthel}, Peter and {Clements}, David L. and {Cohen}, Martin and {Doi}, Yasuo and {Enya}, Keigo and {Figueredo}, Elysandra and {Fujishiro}, Naofumi and {Fujiwara}, Hideaki and {Fujiwara}, Mikio and {Garcia-Lario}, Pedro and {Goto}, Tomotsugu and {Hasegawa}, Sunao and {Hibi}, Yasunori and {Hirao}, Takanori and {Hiromoto}, Norihisa and {Hong}, Seung Soo and {Imai}, Koji and {Ishigaki}, Miho and {Ishiguro}, Masateru and {Ishihara}, Daisuke and {Ita}, Yoshifusa and {Jeong}, Woong-Seob and {Jeong}, Kyung Sook and {Kaneda}, Hidehiro and {Kataza}, Hirokazu and {Kawada}, Mitsunobu and {Kawai}, Toshihide and {Kawamura}, Akiko and {Kessler}, Martin F. and {Kester}, Do and {Kii}, Tsuneo and {Kim}, Dong Chan and {Kim}, Woojung and {Kobayashi}, Hisato and {Koo}, Bon Chul and {Kwon}, Suk Minn and {Lee}, Hyung Mok and {Lorente}, Rosario and {Makiuti}, Sin'itirou and {Matsuhara}, Hideo and {Matsumoto}, Toshio and {Matsuo}, Hiroshi and {Matsuura}, Shuji and {M{\"U}ller}, Thomas G. and {Murakami}, Noriko and {Nagata}, Hirohisa and {Nakagawa}, Takao and {Naoi}, Takahiro and {Narita}, Masanao and {Noda}, Manabu and {Oh}, Sang Hoon and {Ohnishi}, Akira and {Ohyama}, Youichi and {Okada}, Yoko and {Okuda}, Haruyuki and {Oliver}, Sebastian and {Onaka}, Takashi and {Ootsubo}, Takafumi and {Oyabu}, Shinki and {Pak}, Soojong and {Park}, Yong-Sun and {Pearson}, Chris P. and {Rowan-Robinson}, Michael and {Saito}, Toshinobu and {Sakon}, Itsuki and {Salama}, Alberto and {Sato}, Shinji and {Savage}, Richard S. and {Serjeant}, Stephen and {Shibai}, Hiroshi and {Shirahata}, Mai and {Sohn}, Jungjoo and {Suzuki}, Toyoaki and {Takagi}, Toshinobu and {Takahashi}, Hidenori and {Tanab{\'E}}, Toshihiko and {Takeuchi}, Tsutomu T. and {Takita}, Satoshi and {Thomson}, Matthew and {Uemizu}, Kazunori and {Ueno}, Munetaka and {Usui}, Fumihiko and {Verdugo}, Eva and {Wada}, Takehiko and {Wang}, Lingyu and {Watabe}, Toyoki and {Watarai}, Hidenori and {White}, Glenn J. and {Yamamura}, Issei and {Yamauchi}, Chisato and {Yasuda}, Akiko},
        title = "{The Infrared Astronomical Mission AKARI*}",
      journal = {\pasj},
         year = 2007,
        month = oct,
       volume = {59},
        pages = {S369-S376},
          doi = {10.1093/pasj/59.sp2.S369},
archivePrefix = {arXiv},
       eprint = {0708.1796},
 primaryClass = {astro-ph},
       adsurl = {https://ui.adsabs.harvard.edu/abs/2007PASJ...59S.369M}
}

@ARTICLE{Nagao+01a,
       author = {{Nagao}, Tohru and {Murayama}, Takashi and {Taniguchi}, Yoshiaki},
        title = "{Seyfert-Type Dependences of Narrow Emission-Line Ratios and Physical Properties of High-Ionization Nuclear Emission-Line Regions in Seyfert Galaxies}",
      journal = {\pasj},
         year = 2001,
        month = aug,
       volume = {53},
       number = {4},
        pages = {629-645},
          doi = {10.1093/pasj/53.4.629},
archivePrefix = {arXiv},
       eprint = {astro-ph/0107025},
 primaryClass = {astro-ph},
       adsurl = {https://ui.adsabs.harvard.edu/abs/2001PASJ...53..629N}
}

@ARTICLE{Nagao+01b,
       author = {{Nagao}, Tohru and {Murayama}, Takashi and {Taniguchi}, Yoshiaki},
        title = "{Where is the[O III] {\ensuremath{\lambda}}4363 Emitting Region in Active Galactic Nuclei?}",
      journal = {\apj},
         year = 2001,
        month = mar,
       volume = {549},
       number = {1},
        pages = {155-171},
          doi = {10.1086/319062},
archivePrefix = {arXiv},
       eprint = {astro-ph/0011105},
 primaryClass = {astro-ph},
       adsurl = {https://ui.adsabs.harvard.edu/abs/2001ApJ...549..155N}
}

@ARTICLE{Nagao+01c,
       author = {{Nagao}, Tohru and {Murayama}, Takashi and {Taniguchi}, Yoshiaki},
        title = "{The Narrow-Line Region of Seyfert Galaxies: Narrow-Line Seyfert 1 Galaxies versus Broad-Line Seyfert 1 Galaxies}",
      journal = {\apj},
         year = 2001,
        month = jan,
       volume = {546},
       number = {2},
        pages = {744-758},
          doi = {10.1086/318300},
archivePrefix = {arXiv},
       eprint = {astro-ph/0008006},
 primaryClass = {astro-ph},
       adsurl = {https://ui.adsabs.harvard.edu/abs/2001ApJ...546..744N}
}

@ARTICLE{Narayanan+10,
       author = {{Narayanan}, Desika and {Dey}, Arjun and {Hayward}, Christopher C. and {Cox}, Thomas J. and {Bussmann}, R. Shane and {Brodwin}, Mark and {Jonsson}, Patrik and {Hopkins}, Philip F. and {Groves}, Brent and {Younger}, Joshua D. and {Hernquist}, Lars},
        title = "{A physical model for z \raisebox{-0.5ex}\textasciitilde 2 dust-obscured galaxies}",
      journal = {\mnras},
         year = 2010,
        month = sep,
       volume = {407},
       number = {3},
        pages = {1701-1720},
          doi = {10.1111/j.1365-2966.2010.16997.x},
archivePrefix = {arXiv},
       eprint = {0910.2234},
 primaryClass = {astro-ph.CO},
       adsurl = {https://ui.adsabs.harvard.edu/abs/2010MNRAS.407.1701N}
}

@ARTICLE{Nenkova+08a,
       author = {{Nenkova}, Maia and {Sirocky}, Matthew M. and {Ivezi{\'c}}, {\v{Z}}eljko and {Elitzur}, Moshe},
        title = "{AGN Dusty Tori. I. Handling of Clumpy Media}",
      journal = {\apj},
         year = 2008,
        month = sep,
       volume = {685},
       number = {1},
        pages = {147-159},
          doi = {10.1086/590482},
archivePrefix = {arXiv},
       eprint = {0806.0511},
 primaryClass = {astro-ph},
       adsurl = {https://ui.adsabs.harvard.edu/abs/2008ApJ...685..147N}
}

@ARTICLE{Nenkova+08b,
       author = {{Nenkova}, Maia and {Sirocky}, Matthew M. and {Nikutta}, Robert and {Ivezi{\'c}}, {\v{Z}}eljko and {Elitzur}, Moshe},
        title = "{AGN Dusty Tori. II. Observational Implications of Clumpiness}",
      journal = {\apj},
         year = 2008,
        month = sep,
       volume = {685},
       number = {1},
        pages = {160-180},
          doi = {10.1086/590483},
archivePrefix = {arXiv},
       eprint = {0806.0512},
 primaryClass = {astro-ph},
       adsurl = {https://ui.adsabs.harvard.edu/abs/2008ApJ...685..160N}
}

@ARTICLE{Nobo+19,
       author = {{Noboriguchi}, Akatoki and {Nagao}, Tohru and {Toba}, Yoshiki and {Niida}, Mana and {Kajisawa}, Masaru and {Onoue}, Masafusa and {Matsuoka}, Yoshiki and {Yamashita}, Takuji and {Chang}, Yu-Yen and {Kawaguchi}, Toshihiro and {Komiyama}, Yutaka and {Nobuhara}, Kodai and {Terashima}, Yuichi and {Ueda}, Yoshihiro},
        title = "{Optical Properties of Infrared-bright Dust-obscured Galaxies Viewed with Subaru Hyper Suprime-Cam}",
      journal = {\apj},
         year = 2019,
        month = may,
       volume = {876},
       number = {2},
          eid = {132},
        pages = {132},
          doi = {10.3847/1538-4357/ab1754},
archivePrefix = {arXiv},
       eprint = {1803.09951},
 primaryClass = {astro-ph.GA},
       adsurl = {https://ui.adsabs.harvard.edu/abs/2019ApJ...876..132N}
}

@ARTICLE{Perrotta+19,
       author = {{Perrotta}, S. and {Hamann}, F. and {Zakamska}, N.~L. and {Alexandroff}, R.~M. and {Rupke}, D. and {Wylezalek}, D.},
        title = "{ERQs are the BOSS of quasar samples: the highest velocity [O III] quasar outflows}",
      journal = {\mnras},
         year = 2019,
        month = sep,
       volume = {488},
       number = {3},
        pages = {4126-4148},
          doi = {10.1093/mnras/stz1993},
archivePrefix = {arXiv},
       eprint = {1906.00980},
 primaryClass = {astro-ph.GA},
       adsurl = {https://ui.adsabs.harvard.edu/abs/2019MNRAS.488.4126P}
}

@ARTICLE{Pilbratt+10,
       author = {{Pilbratt}, G.~L. and {Riedinger}, J.~R. and {Passvogel}, T. and {Crone}, G. and {Doyle}, D. and {Gageur}, U. and {Heras}, A.~M. and {Jewell}, C. and {Metcalfe}, L. and {Ott}, S. and {Schmidt}, M.},
        title = "{Herschel Space Observatory. An ESA facility for far-infrared and submillimetre astronomy}",
      journal = {\aap},
         year = 2010,
        month = jul,
       volume = {518},
          eid = {L1},
        pages = {L1},
          doi = {10.1051/0004-6361/201014759},
archivePrefix = {arXiv},
       eprint = {1005.5331},
 primaryClass = {astro-ph.IM},
       adsurl = {https://ui.adsabs.harvard.edu/abs/2010A&A...518L...1P}
}

@ARTICLE{Ramos+25,
       author = {{Ramos Almeida}, C. and {Garc{\'\i}a-Bernete}, I. and {Pereira-Santaella}, M. and {Speranza}, G. and {Maiolino}, R. and {Ji}, X. and {Audibert}, A. and {Cezar}, P.~H. and {Acosta-Pulido}, J.~A. and {Alonso-Herrero}, A. and {Garc{\'\i}a-Burillo}, S. and {Gonz{\'a}lez-Mart{\'\i}n}, O. and {Rigopoulou}, D. and {Tadhunter}, C.~N. and {Labiano}, A. and {Levenson}, N.~A. and {Donnan}, F.~R.},
        title = "{JWST MIRI reveals the diversity of nuclear mid-infrared spectra of nearby type 2 quasars}",
      journal = {\aap},
         year = 2025,
        month = jun,
       volume = {698},
          eid = {A194},
        pages = {A194},
          doi = {10.1051/0004-6361/202453549},
archivePrefix = {arXiv},
       eprint = {2504.01595},
 primaryClass = {astro-ph.GA},
       adsurl = {https://ui.adsabs.harvard.edu/abs/2025A&A...698A.194R}
}

@ARTICLE{Rod+13,
       author = {{Rodr{\'\i}guez Zaur{\'\i}n}, J. and {Tadhunter}, C.~N. and {Rose}, M. and {Holt}, J.},
        title = "{The importance of warm, AGN-driven outflows in the nuclear regions of nearby ULIRGs}",
      journal = {\mnras},
         year = 2013,
        month = jun,
       volume = {432},
       number = {1},
        pages = {138-166},
          doi = {10.1093/mnras/stt423},
archivePrefix = {arXiv},
       eprint = {1303.1400},
 primaryClass = {astro-ph.CO},
       adsurl = {https://ui.adsabs.harvard.edu/abs/2013MNRAS.432..138R}
}

@ARTICLE{Rose+18,
       author = {{Rose}, Marvin and {Tadhunter}, Clive and {Ramos Almeida}, Cristina and {Rodr{\'\i}guez Zaur{\'\i}n}, Javier and {Santoro}, Francesco and {Spence}, Robert},
        title = "{Quantifying the AGN-driven outflows in ULIRGs (QUADROS) - I: VLT/Xshooter observations of nine nearby objects}",
      journal = {\mnras},
         year = 2018,
        month = feb,
       volume = {474},
       number = {1},
        pages = {128-156},
          doi = {10.1093/mnras/stx2590},
archivePrefix = {arXiv},
       eprint = {1710.06600},
 primaryClass = {astro-ph.GA},
       adsurl = {https://ui.adsabs.harvard.edu/abs/2018MNRAS.474..128R}
}

@ARTICLE{Ross+15,
       author = {{Ross}, Nicholas P. and {Hamann}, Fred and {Zakamska}, Nadia L. and {Richards}, Gordon T. and {Villforth}, Carolin and {Strauss}, Michael A. and {Greene}, Jenny E. and {Alexandroff}, Rachael and {Brandt}, W. Niel and {Liu}, Guilin and {Myers}, Adam D. and {P{\^a}ris}, Isabelle and {Schneider}, Donald P.},
        title = "{Extremely red quasars from SDSS, BOSS and WISE: classification of optical spectra}",
      journal = {\mnras},
         year = 2015,
        month = nov,
       volume = {453},
       number = {4},
        pages = {3932-3952},
          doi = {10.1093/mnras/stv1710},
archivePrefix = {arXiv},
       eprint = {1405.1047},
 primaryClass = {astro-ph.GA},
       adsurl = {https://ui.adsabs.harvard.edu/abs/2015MNRAS.453.3932R}
}

@ARTICLE{Sanders+88,
       author = {{Sanders}, D.~B. and {Soifer}, B.~T. and {Elias}, J.~H. and {Madore}, B.~F. and {Matthews}, K. and {Neugebauer}, G. and {Scoville}, N.~Z.},
        title = "{Ultraluminous Infrared Galaxies and the Origin of Quasars}",
      journal = {\apj},
         year = 1988,
        month = feb,
       volume = {325},
        pages = {74},
          doi = {10.1086/165983},
       adsurl = {https://ui.adsabs.harvard.edu/abs/1988ApJ...325...74S}
}

@ARTICLE{Skrutskie+06,
       author = {{Skrutskie}, M.~F. and {Cutri}, R.~M. and {Stiening}, R. and {Weinberg}, M.~D. and {Schneider}, S. and {Carpenter}, J.~M. and {Beichman}, C. and {Capps}, R. and {Chester}, T. and {Elias}, J. and {Huchra}, J. and {Liebert}, J. and {Lonsdale}, C. and {Monet}, D.~G. and {Price}, S. and {Seitzer}, P. and {Jarrett}, T. and {Kirkpatrick}, J.~D. and {Gizis}, J.~E. and {Howard}, E. and {Evans}, T. and {Fowler}, J. and {Fullmer}, L. and {Hurt}, R. and {Light}, R. and {Kopan}, E.~L. and {Marsh}, K.~A. and {McCallon}, H.~L. and {Tam}, R. and {Van Dyk}, S. and {Wheelock}, S.},
        title = "{The Two Micron All Sky Survey (2MASS)}",
      journal = {\aj},
         year = 2006,
        month = feb,
       volume = {131},
       number = {2},
        pages = {1163-1183},
          doi = {10.1086/498708},
       adsurl = {https://ui.adsabs.harvard.edu/abs/2006AJ....131.1163S}
}

@ARTICLE{Spence+18,
       author = {{Spence}, R.~A.~W. and {Tadhunter}, C.~N. and {Rose}, M. and {Rodr{\'\i}guez Zaur{\'\i}n}, J.},
        title = "{Quantifying the AGN-driven outflows in ULIRGs (QUADROS) III: measurements of the radii and kinetic powers of eight near-nuclear outflows}",
      journal = {\mnras},
         year = 2018,
        month = aug,
       volume = {478},
       number = {2},
        pages = {2438-2460},
          doi = {10.1093/mnras/sty1046},
archivePrefix = {arXiv},
       eprint = {1805.02647},
 primaryClass = {astro-ph.GA},
       adsurl = {https://ui.adsabs.harvard.edu/abs/2018MNRAS.478.2438S}
}

@ARTICLE{Storey+00,
       author = {{Storey}, P.~J. and {Zeippen}, C.~J.},
        title = "{Theoretical values for the [OIII] 5007/4959 line-intensity ratio and homologous cases}",
      journal = {\mnras},
         year = 2000,
        month = mar,
       volume = {312},
       number = {4},
        pages = {813-816},
          doi = {10.1046/j.1365-8711.2000.03184.x},
       adsurl = {https://ui.adsabs.harvard.edu/abs/2000MNRAS.312..813S}
}

@misc{STScI,
  doi = {10.17909/T9H59D},
  url = {http://archive.stsci.edu/doi/resolve/resolve.html?doi=10.17909/T9H59D},
  author = {{STScI}},
  title = {GALEX/MCAT},
  publisher = {STScI/MAST},
  year = {2013}
}

@ARTICLE{Suleiman+22,
       author = {{Suleiman}, Nofoz and {Noboriguchi}, Akatoki and {Toba}, Yoshiki and {Bal{\'a}zs}, Lajos G. and {Burgarella}, Denis and {Kov{\'a}cs}, Timea and {Marton}, G{\'a}bor and {Talafha}, Mohammed and {Frey}, S{\'a}ndor and {T{\'o}th}, L. Viktor},
        title = "{The statistical properties of 28 IR-bright dust-obscured galaxies and SED modelling using CIGALE}",
      journal = {\pasj},
         year = 2022,
        month = oct,
       volume = {74},
       number = {5},
        pages = {1157-1185},
          doi = {10.1093/pasj/psac061},
       adsurl = {https://ui.adsabs.harvard.edu/abs/2022PASJ...74.1157S}
}

@ARTICLE{Tadhunter+18,
       author = {{Tadhunter}, C. and {Rodr{\'\i}guez Zaur{\'\i}n}, J. and {Rose}, M. and {Spence}, R.~A.~W. and {Batcheldor}, D. and {Berg}, M.~A. and {Ramos Almeida}, C. and {Spoon}, H.~W.~W. and {Sparks}, W. and {Chiaberge}, M.},
        title = "{Quantifying the AGN-driven outflows in ULIRGs (QUADROS) - II. Evidence for compact outflow regions from HST [O III] imaging observations}",
      journal = {\mnras},
         year = 2018,
        month = aug,
       volume = {478},
       number = {2},
        pages = {1558-1569},
          doi = {10.1093/mnras/sty1064},
archivePrefix = {arXiv},
       eprint = {1805.00514},
 primaryClass = {astro-ph.GA},
       adsurl = {https://ui.adsabs.harvard.edu/abs/2018MNRAS.478.1558T}
}

@ARTICLE{Tadhunter+19,
       author = {{Tadhunter}, C. and {Holden}, L. and {Ramos Almeida}, C. and {Batcheldor}, D.},
        title = "{Quantifying the AGN-driven outflows in ULIRGs (QUADROS) IV: HST/STIS spectroscopy of the sub-kpc warm outflow in F14394+5332}",
      journal = {\mnras},
         year = 2019,
        month = sep,
       volume = {488},
       number = {2},
        pages = {1813-1821},
          doi = {10.1093/mnras/stz1755},
archivePrefix = {arXiv},
       eprint = {1908.03104},
 primaryClass = {astro-ph.GA},
       adsurl = {https://ui.adsabs.harvard.edu/abs/2019MNRAS.488.1813T}
}

@ARTICLE{Tanimoto+19,
       author = {{Tanimoto}, Atsushi and {Ueda}, Yoshihiro and {Odaka}, Hirokazu and {Kawaguchi}, Toshihiro and {Fukazawa}, Yasushi and {Kawamuro}, Taiki},
        title = "{XCLUMPY: X-Ray Spectral Model from Clumpy Torus and Its Application to the Circinus Galaxy}",
      journal = {\apj},
         year = 2019,
        month = jun,
       volume = {877},
       number = {2},
          eid = {95},
        pages = {95},
          doi = {10.3847/1538-4357/ab1b20},
archivePrefix = {arXiv},
       eprint = {1904.08945},
 primaryClass = {astro-ph.HE},
       adsurl = {https://ui.adsabs.harvard.edu/abs/2019ApJ...877...95T}
}

@ARTICLE{Toba+16,
       author = {{Toba}, Y. and {Nagao}, T.},
        title = "{Search for Hyperluminous Infrared Dust-obscured Galaxies Selected with WISE and SDSS}",
      journal = {\apj},
         year = 2016,
        month = mar,
       volume = {820},
       number = {1},
          eid = {46},
        pages = {46},
          doi = {10.3847/0004-637X/820/1/46},
archivePrefix = {arXiv},
       eprint = {1602.07870},
 primaryClass = {astro-ph.GA},
       adsurl = {https://ui.adsabs.harvard.edu/abs/2016ApJ...820...46T}
}

@ARTICLE{Toba+17,
       author = {{Toba}, Yoshiki and {Bae}, Hyun-Jin and {Nagao}, Tohru and {Woo}, Jong-Hak and {Wang}, Wei-Hao and {Wagner}, Alexander Y. and {Sun}, Ai-Lei and {Chang}, Yu-Yen},
        title = "{Ionized Gas Outflows in Infrared-bright Dust-obscured Galaxies Selected with WISE  and SDSS}",
      journal = {\apj},
         year = 2017,
        month = dec,
       volume = {850},
       number = {2},
          eid = {140},
        pages = {140},
          doi = {10.3847/1538-4357/aa918a},
archivePrefix = {arXiv},
       eprint = {1710.02525},
 primaryClass = {astro-ph.GA},
       adsurl = {https://ui.adsabs.harvard.edu/abs/2017ApJ...850..140T}
}

@ARTICLE{Toba+18,
       author = {{Toba}, Yoshiki and {Ueda}, Junko and {Lim}, Chen-Fatt and {Wang}, Wei-Hao and {Nagao}, Tohru and {Chang}, Yu-Yen and {Saito}, Toshiki and {Kawabe}, Ryohei},
        title = "{Discovery of an Extremely Luminous Dust-obscured Galaxy Observed with SDSS, WISE, JCMT, and SMA}",
      journal = {\apj},
         year = 2018,
        month = apr,
       volume = {857},
       number = {1},
          eid = {31},
        pages = {31},
          doi = {10.3847/1538-4357/aab3cf},
archivePrefix = {arXiv},
       eprint = {1803.00177},
 primaryClass = {astro-ph.GA},
       adsurl = {https://ui.adsabs.harvard.edu/abs/2018ApJ...857...31T}
}

@ARTICLE{Toba+24,
       author = {{Toba}, Yoshiki and {Hashiguchi}, Aoi and {Ota}, Naomi and {Oguri}, Masamune and {Okabe}, Nobuhiro and {Ueda}, Yoshihiro and {Imanishi}, Masatoshi and {Nishizawa}, Atsushi J. and {Goto}, Tomotsugu and {Hsieh}, Bau-Ching and {Kondo}, Marie and {Koyama}, Shuhei and {Lee}, Kianhong and {Mitsuishi}, Ikuyuki and {Nagao}, Tohru and {Oogi}, Taira and {Sakuta}, Koki and {Schramm}, Malte and {Yanagawa}, Anri and {Yoshimoto}, Anje},
        title = "{Active Galactic Nucleus Properties of {\ensuremath{\sim}}1 Million Member Galaxies of Galaxy Groups and Clusters at z < 1.4 Based on the Subaru Hyper Suprime-Cam Survey}",
      journal = {\apj},
         year = 2024,
        month = may,
       volume = {967},
       number = {1},
          eid = {65},
        pages = {65},
          doi = {10.3847/1538-4357/ad32c6},
archivePrefix = {arXiv},
       eprint = {2402.11188},
 primaryClass = {astro-ph.GA},
       adsurl = {https://ui.adsabs.harvard.edu/abs/2024ApJ...967...65T}
}

@ARTICLE{Wada+12,
       author = {{Wada}, Keiichi},
        title = "{Radiation-driven Fountain and Origin of Torus around Active Galactic Nuclei}",
      journal = {\apj},
         year = 2012,
        month = oct,
       volume = {758},
       number = {1},
          eid = {66},
        pages = {66},
          doi = {10.1088/0004-637X/758/1/66},
archivePrefix = {arXiv},
       eprint = {1208.5272},
 primaryClass = {astro-ph.GA},
       adsurl = {https://ui.adsabs.harvard.edu/abs/2012ApJ...758...66W}
}
\bibliographystyle{aasjournal}

\end{document}